\documentclass[prb,twocolumn,showpacs,preprintnumbers,amsmath,amssymb]{revtex4}% Physical Review B

\usepackage{graphicx}% Include figure files
\graphicspath{{fig/}}
\usepackage{dcolumn}% Align table columns on decimal point\varphi 
\usepackage{bm}% bold math
\usepackage{amsmath,amssymb}
\usepackage[T1]{fontenc}
\usepackage{textcomp}
\usepackage{color}
\begin{document}

\title{
Magnonic quantum Hall effect  and Wiedemann-Franz law
}

\author{Kouki Nakata, Jelena Klinovaja, and Daniel Loss}

\affiliation{Department of Physics, University of Basel, Klingelbergstrasse 82, CH-4056 Basel, Switzerland   
}

\date{\today}

\begin{abstract}
We present a quantum Hall effect of magnons in  two-dimensional clean insulating magnets at finite temperature.
Through the Aharonov-Casher effect, a  magnon moving in an electric field acquires a geometric phase and forms Landau levels in an electric  field gradient of sawtooth form. 
At low temperatures, the lowest energy band being almost flat carries a Chern number associated with a Berry curvature.
Appropriately defining the thermal conductance for bosons, we find that the magnon Hall conductances get quantized and show a universal thermomagnetic behavior, i.e., are independent of materials, and obey a Wiedemann-Franz law for magnon transport. 
We consider magnons with quadratic and linear (Dirac-like) dispersions.
Finally, we show that our predictions are within experimental reach for ferromagnets and skyrmion lattices with current device and measurement techniques.
\end{abstract}

\pacs{75.30.Ds, 73.43.-f, 77.55.Nv, 03.65.Pm}

\maketitle

%%%%%%%%%%%%%%%%%
\section{Introduction}
\label{sec:Intro}
%%%%%%%%%%%%%%%%%
Magnons \cite{BlochMagnon,HP,DanielMagnetismBook}, the quantized version of spin waves, are low-energy collective excitations of  coupled localized spins and play the role of an elemental magnetic carrier in a wide range of insulating magnets.
Due to the intrinsic bosonic nature, magnons can form a macroscopic coherent state by quasi-equilibrium condensation \cite{demokritov,ultrahot,SergaBEC,MagnonSupercurrent,bunkov} 
and propagate \cite{spinwave} spin information over distances of several millimeters, much further than what is typically possible when using spin-polarized conduction electrons in metals.
Such fascinating properties have attracted considerable interest in magnon spintronics, dubbed magnonics \cite{MagnonSpintronics,magnonics,spincalreview,ReviewPRapply},
aimed at utilizing magnons,  instead of charge, as a carrier of information in units of the Bohr magneton $\mu_{\rm{B}}$.

For this purpose, it is of fundamental interest to develop a better  understanding of magnon transport in magnetic insulators\cite{magnon2,magnonWF,KKPD,KPD,ReviewMagnon,KevinPRL,KevinHallEffect,KevinHighliht,TserkovnyakNatNano,NOTEreview,UtrechtGirl,YacobyChemical}.
%%%%%%%%%%%%%%%%%%%%%%%%
Analogous to the Aharonov-Bohm \cite{bohm,LossPersistent,LossPersistent2} (AB)  effect of charged particles  in magnetic fields, a magnetic dipole  moving  in electric fields acquires a geometric phase called the Aharonov-Casher \cite{casher,Mignani} (AC) phase.
Because magnons have a magnetic dipole moment, this AC effect gives a handle to electrically control magnon transport \cite{magnon2,katsura2}.
Zhang \textit{et al.} \cite{ACspinwave} have indeed experimentally observed the AC effect in magnon systems, and a method to electromagnetically control magnon transport in condensation \cite{MagnonSupercurrent}, magnon Josephson effects, and persistent quantized magnon current, have been proposed recently~\cite{KKPD} using the AC effect.
%%%%%%%%%%%%%%%%%%%%%%%%%%
As well as such electromagnetic aspects, also thermomagnetic control has been rapidly developed.
Making use of the thermal Hall effect \cite{onose,katsura,Matsumoto,Matsumoto2}, Onose \textit{et al.} \cite{onose} have experimentally realized the thermomagnetic control of magnon transport in  pyrochlore structured magnets and recently, a universal thermomagnetic relation, the Wiedemann-Franz \cite{WFgermany} (WF) law, 
for magnon transport has been theoretically established~\cite{magnonWF}; at temperatures sufficiently lower than the energy gap, provided by the Zeeman energy,
the ratio between thermal and magnetic magnon-conductances becomes  linear in temperature with a universal proportionality constant.

\begin{figure}[t]
\begin{center}
\includegraphics[width=7cm,clip]{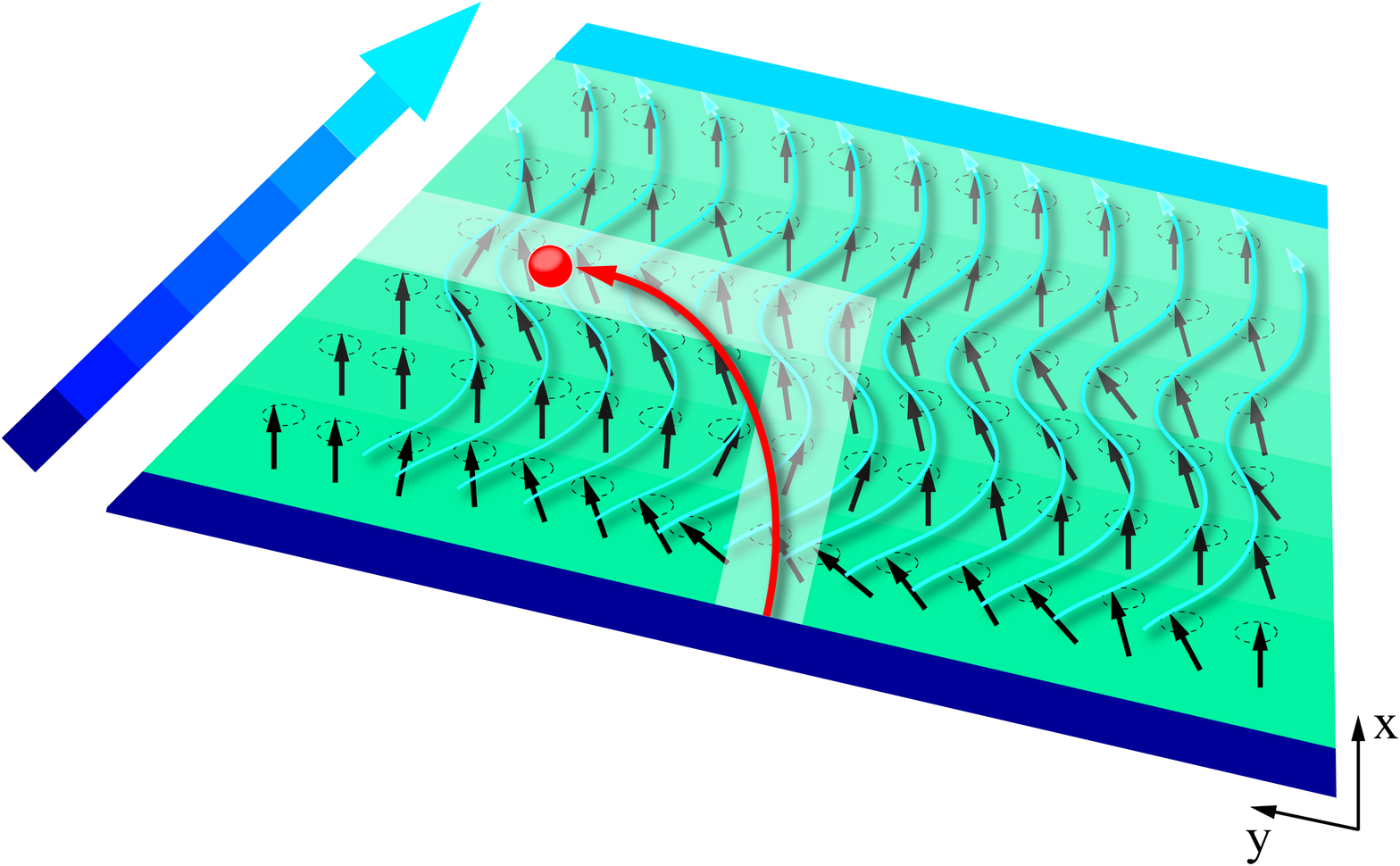}
\caption{(Color online)
Schematic representation of magnonic Hall effects in a two-dimensional clean insulating magnet where magnetic field or temperature gradients along $x$ direction produce a transverse Hall current of magnon (red dot) along $y$ direction. Landau levels and cyclotron motion of the magnons can be induced by extrinsic electric field gradients or intrinsic spin orbit interactions, giving rise to quantized Hall conductances for magnon and thermal currents. The ratio of these Hall conductances satisfies a Wiedemann-Franz law at low temperatures.
\label{fig:MagnonicSystem} }
\end{center}
\end{figure}

In this work, providing a topological description \cite{NiuBerry,Kohmoto,TKNN} of the classical magnon Hall effect induced by the AC phase, which was proposed in Ref. [\onlinecite{magnon2}], we develop it further into  a quantum Hall effect (QHE) of magnons and find a universal thermomagnetic behavior.
The mathematical structure of magnons in the presence of an AC phase is identical to that of electrons with an AB phase, which allows us to use the topological formulations \cite{NiuBerry,Kohmoto,TKNN} of the QHE in terms of Chern numbers (i.e., topological invariant) and apply them to our case.
%%%%%%%%%%%%%%%%%%%%%
Indeed, a QHE of spin currents \cite{LeggettRevHe} in Helium-3 has been proposed by Volovik and Yakovenko \cite{VolovikQHE}, and similarly in disordered \cite{NiuPhaseTwist} chiral spin liquids at zero temperature by Haldane and Arovas \cite{Haldane2}.
Here, we  focus on a general clean insulating magnet at finite temperatures and propose a QHE of magnons in Landau quantization induced by strong electric field gradients or by an intrinsic spin texture such as two-dimensional skyrmion lattices~\cite{KevinHallEffect}.
We show that the ratio of the magnon and thermal quantum Hall conductance becomes universal and satisfies the Wiedemann-Franz law, provided the proper definition of the thermal Hall conductance is used which includes off-diagonal Onsager coefficients.
We present numerics for the integer QHE in the presence of a periodically extended gradient field of sawtooth form. 
We find that for this case the edge states remain intact even if the period of the sawtooth is much smaller than the electric length characterizing the Landau orbit size.
We consider magnons with quadratic and linear (Dirac-like) dispersions and discuss the differences of the corresponding Landau levels.

At sufficiently low temperatures, effects of magnon-magnon interactions and magnon-phonons become \cite{magnonWF,adachiphonon} negligibly small.
We then focus on such low temperatures throughout this paper and assume noninteracting \footnote{Within the mean-field treatment, magnon-magnon interactions indeed reduce \cite{magnonWF} to an effective magnetic field and the results qualitatively remain the same. 
A certain class of QHEs in systems with interacting bosons can be found in Refs.~[\onlinecite{SenthilLevin}] and [\onlinecite{SPTcoldatom}]. See Refs. [\onlinecite{RShindou,RShindou2,RShindou3}] for chiral edge states in systems with  dipolar interactions and for the bulk-edge correspondence.}
magnons.

This paper is organized as follows.
In Sec. \ref{sec:LL} we introduce the model system for magnons with a quadratic dispersion, applicable to wide range of insulating magnetic lattices (i.e., crystals),
and find the Landau quantization through the AC effects. 
In Sec. \ref{sec:QHE}, analyzing the resulting Hall conductances, we discuss the condition for the QHE that is characterized by a Chern number associated with the Berry curvature and derive the WF law in the quantum Hall system.
In Sec. \ref{sec:experiments}, we give some concrete estimates for experimental candidate materials.
Finally, we summarize and give some conclusions in Sec. \ref{sec:sum}.
Technical details are deferred to Appendices.

\begin{figure}[t]
\begin{center}
\includegraphics[width=6cm,clip]{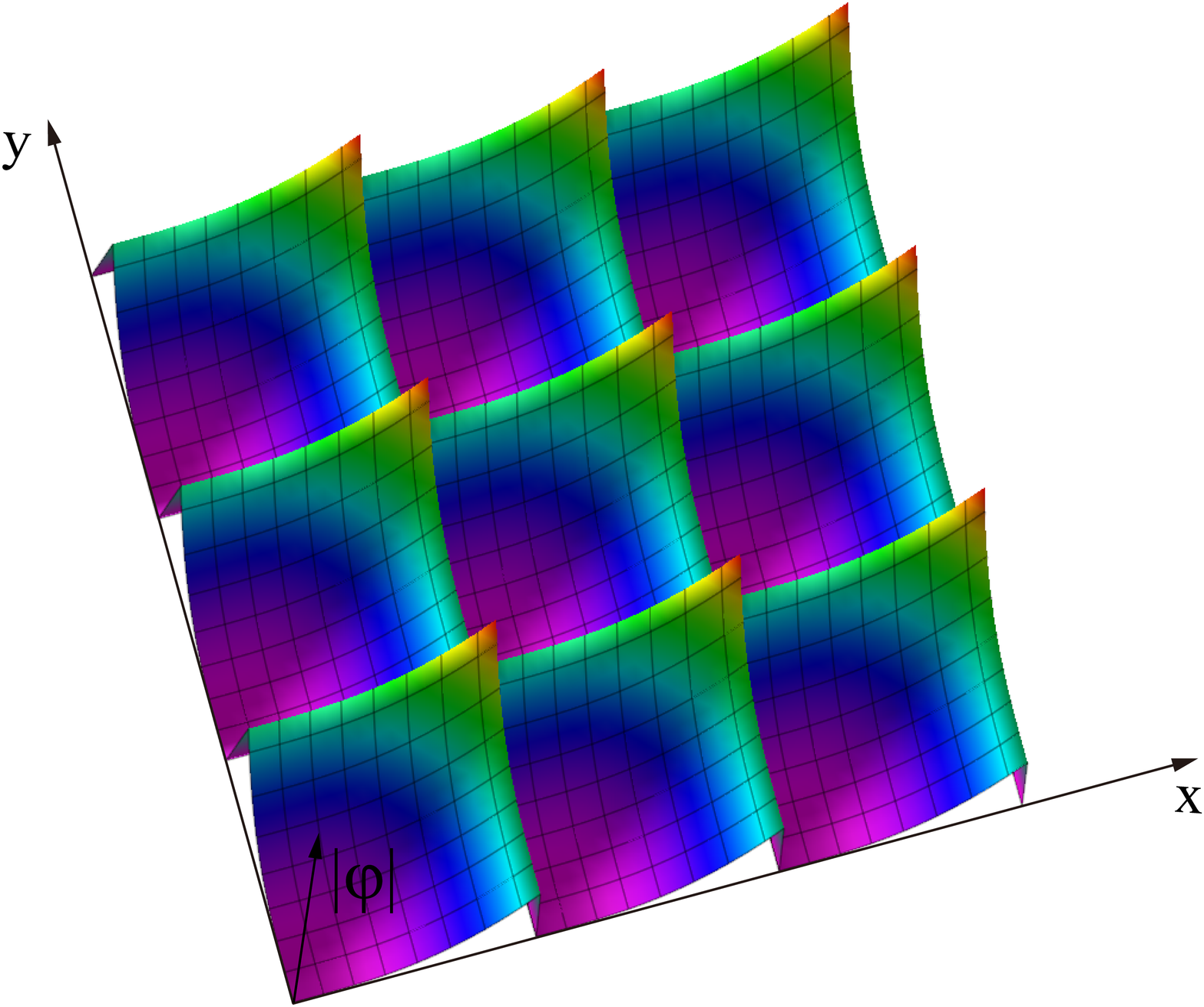}
\caption{(Color online)
Schematic representation of a skew-harmonic electromagnetic scalar potential $\varphi $ periodically arranged.
Through the AC effect, the magnon ($\mu _{\rm{m}}$) moving in the potential experiences an electric vector potential ${\mathbf{A}}_{\rm{m}}$ analogous to the symmetric gauge giving rise to cyclotron motion of the magnon corresponding to Landau levels.
Such a skew-harmonic potential may be realized  by periodically arranging STM tips.
\label{fig:PeriodicPotential} }
\end{center}
\end{figure}

%%%%%%%%%%%%%%%%%%%%%%%%%%%
\section{Landau levels}
\label{sec:LL}
%%%%%%%%%%%%%%%%%%%%%%%%%%%
We consider a two-dimensional  clean ferromagnet (see Fig. \ref{fig:MagnonicSystem}) embedded in the $xy$ plane in the presence of an electric field ${\mathbf{E}}$ which couples to the magnetic dipole $ g \mu _{\rm{B}}{\mathbf{e}}_z$ of the spins through the AC effect \cite{Mook2,Lifa,katsura2,ACspinwave}, where $g$ is the $g$-factor and $\mu_{\rm{B}}$ the Bohr magneton. The magnet  is described by an anisotropic Heisenberg spin Hamiltonian given by~\cite{magnon2}
\begin{eqnarray}
{\cal{H}}  = -  \sum_{\langle i j \rangle} J_{ij}  [\frac{1}{2}(S_i ^{+}S_j^{-} {\rm{e}}^{ i \theta _{ij}}+S_i ^{-}S_j^{+} {\rm{e}}^{- i \theta _{ij}}) +S_i ^{z}S_j^{z}],
\label{eqn:SpinHamiltonian} 
\end{eqnarray}
where $J_{ij} >0$ is the exchange interaction between the localized spins on the nearest neighboring sites $\langle i j \rangle$, $S_j ^{\pm }= S_j^x \pm i S_j^y $. Here, we allow for spatial anisotropy such that if the exchange bond between site $i$ and $j$ is along $x$ direction, the exchange interaction is given by $J_x$, 
while along $y$ direction by $J_y$. The exchange interaction in spin space is assumed to be isotropic.
Furthermore, 
$\theta _{ij} = (g \mu _{\rm{B}}/\hbar c^2) \int_{{\mathbf{x}}_i}^{{\mathbf{x}}_j} d  {\mathbf{r}} \cdot  ({\mathbf{E}}\times {\mathbf{e}}_z) $ 
is the AC phase \cite{Mook2,Lifa,katsura2,ACspinwave} which the magnetic dipole moment associated with the spin along $z$ acquires when it hops between neighboring sites.
By using the Holstein-Primakoff\cite{HP,KPD} transformation, $ S_i^+ = \sqrt{2S}[1-a_i^\dagger a_i / (2S)]^{1/2} a_i$, $S_i^z = S - a_i^\dagger a_i$, to lowest order since we assume large spins $S \gg  1$, Eq. (\ref{eqn:SpinHamiltonian}) can be mapped onto a system of non-interacting \cite{NOTEmmint} magnons: Chargeless bosonic quasi-particles carrying a magnetic dipole moment $g\mu_{\rm{B}}  {\bf e}_z$ along $z$ direction. These magnons are described by annihilation (creation) operators $a_{i}^{(\dagger)}$ which satisfy bosonic commutation relations $[a_{i}, a_{j}^{\dagger }] = \delta_{i,j}$. 
Dropping irrelevant constants, we then get the hopping Hamiltonian form for the magnons,
\begin{eqnarray}
 {\cal{H}}_{\rm{m}}  =   - \sum_{\langle i j \rangle} t_{ij} {\rm{e}}^{ i \theta _{ij}}
(a_{i} a_{j}^\dagger + {\rm{H.c.}}),
 \label{magnon_hopping}  
\end{eqnarray}
where $t_{ij}=J_{ij} S$ is the hopping amplitude.
Going over to the continuum limit, and in the isotropic limit $J_x=J_y=J$, Eq. (\ref{magnon_hopping}) reduces to \cite{magnon2,Mignani}
\begin{eqnarray}
 {\cal{H}}_{\rm{m}}  =    \frac{1}{2m} \Big({\mathbf{p}}  +  \frac{g \mu _{\rm{B}}}{c}{\mathbf{A}}_{\rm{m}} \Big)^2,
\label{HamiltonianLL} 
\end{eqnarray}
where  $m$ is an effective mass of the   magnons defined by $(2m)^{-1} = JSa^2/\hbar ^2$, with $a$ being the isotropic lattice constant,  ${\mathbf{p}}=(p_x, p_y,0)$  the momentum operator of the magnon in the plane, resulting from the  quadratic dispersion of the magnons, and where  we introduced an `electric' vector potential at position ${\mathbf{r}}=(x,y,0)$
\begin{eqnarray}
{\mathbf{A}}_{\rm{m}}({\mathbf{r}})
 = \frac{1}{c} {\mathbf{E}}({\mathbf{r}})\times {\mathbf{e_z}}
\label{magnon_gauge}
\end{eqnarray}
 for the magnons.
The Hamiltonian Eq. (\ref{HamiltonianLL}) is seen to be formally identical to that of a charged particle moving in a magnetic vector potential,
but where now the coupling constant is given by $g \mu _{\rm{B}}$ instead of the charge $e$ and the electric vector potential ${\mathbf{A}}_{\rm{m}}$.
Assuming an electric field of the form ${\mathbf{E}}({\mathbf{r}})= (E_x, E_y, E_z) = {\mathcal{E}}(x/2, y/2,0)$, where 
$ {\mathcal{E}}$ is a constant electric field gradient, this gives the analogue of the symmetric gauge 
${\mathbf{A}}_{\rm{m}}({\mathbf{r}}) = ({\mathcal{E}}/c)(y/2, -x/2,0)$.  
The role of the perpendicular magnetic field in charged systems is played here by the field gradient $ {\mathcal{E}}$, and such an electric field arises from an electric `skew-harmonic' potential, see Fig. \ref{fig:PeriodicPotential}.
This allows us to use the topological formulations \cite{NiuBerry,Kohmoto} of the conventional QHE in terms of Chern numbers and to apply them directly to our case.

\begin{figure}[t]
\begin{center}
\includegraphics[width=8.5cm,clip]{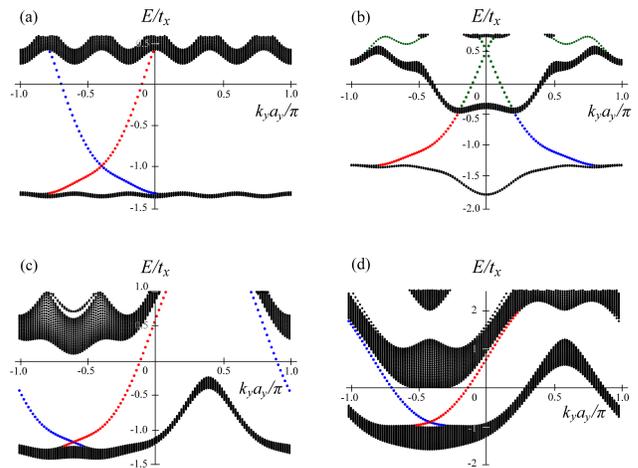}
\caption{(Color online)
Plots of the magnonic band structure, rescaled energy $E/t_x$, in the isotropic limit $t_x=t_y$ as function of the rescaled wavevector $k_y a_y/\pi$ obtained by numerically solving the tight-binding model [Eq. (\ref{TightBinding2})] for the value $\theta _0=2\pi/5$, showing the first and partially the second Landau levels.
The periodicity of the vector potential ${\mathbf{A}}_{\rm{m}q}^{\prime} = ({\mathcal{E}} R_q/c) (0, \{x/R_q \}, 0) $
 is (a) $ q \gg  1$, (b) $q = 6$, (c) $q = 4$, and (d) $q = 3$.
(a): Standard QHE band structure with a well-developed gap  and a chiral  in-gap edge state (one for each edge).
(b): Qualitatively still the same as in (a); with two chiral edge states (green) connecting the second and third Landau level.
(c) and (d):  There are well-defined edge modes for fixed values of $k_y$, but they coexist with extended bulk modes at different momenta. 
Similar plots are obtained for the anisotropic case, see Appendix \ref{sec:aniso}.
\label{fig:Iso_BandSum} }
\end{center}
\end{figure}

Using this analogy, the calculation indeed parallels the one for electrons (see Appendix \ref{sec:MagnonicLLcal} for details).
The energy eigenvalues of $ {\cal{H}}_{\rm{m}}$ become the analog of Landau levels and magnons perform cyclotron motion with frequency 
\begin{eqnarray}
\omega _c  =  \frac{g \mu _{\rm{B}}{\mathcal{E}}}{m c^2}
\label{omega_c} 
\end{eqnarray}
and the electric length $ l_{\rm{{\mathcal{E}}}}$ being defined by
\begin{eqnarray}
 l_{\rm{{\mathcal{E}}}} \equiv  \sqrt{{\hbar c^2}/{g \mu _{\rm{B}}{\mathcal{E}}}},
\label{l_E} 
\end{eqnarray}
which is analogous to the magnetic length \cite{mahan,Ezawa} in charged systems. 
Moreover, it has been established experimentally that magnons satisfy Snell's law at interfaces \cite{Snell_Exp}, in particular implying specular (elastic) reflection at the boundary to vacuum. 
Thus, we can expect that magnons form skipping orbits along the boundary like electrons \cite{HalperinEdge}, giving rise to chiral edge states \cite{RShindou,RShindou2,RShindou3} in the quantum Hall regime, see below  and Fig. \ref{fig:Iso_BandSum}.
%%%%%%%
When, in addition, a uniform magnetic field $B_0$ perpendicular to the $xy$ plane is applied giving rise to a Zeeman energy for the spins, the Landau levels become \cite{ACatom}
\begin{eqnarray}
E_n   = \hbar \omega _c \Big(n +  \frac{1}{2}\Big)  + g \mu _{\rm{B}} B_0  \    \    \      {\rm{for}}     \     \    n \in  {\mathbb{N}}_0.
\label{LL} 
\end{eqnarray}
The energy levels break up into uniformly spaced subbands.
The resulting energy level spacing of such nonrelativistic-like magnons is uniform and does not depend on the principal quantum number $n$ for the Landau level:
 $E_n - E_{n-1}  = \hbar \omega _c$ for $ n \in  {\mathbb{N}}_+$. 
See Appendix \ref{sec:DM} for {\it{Dirac magnons}}~\cite{DiracMagnon} with a linear dispersion.

One of the challenges in above picture is that the electric field gradients ${\mathcal{E}}$ needed for the formation of Landau levels must be very large to reach level spacing that are physically observable. This requirement of large fields can be substantially softened by allowing for a periodic extension of linear field gradients, see Fig. \ref{fig:PeriodicPotential}. In this case, the field gradient is still the same but needs to be generated only over a distance (period) that can be much smaller than the sample dimensions or even the electric length $ l_{\rm{{\mathcal{E}}}}$.

To investigate this we have performed exact numerical diagonalization of the tight-binding Hamiltonian given in Eq. (\ref{magnon_hopping}), as we explain next.
For this it is convenient to work with the analogue of the Landau gauge, ${\mathbf{A}}_{\rm{m}}^{\prime} = ({\mathcal{E}}/c)(0, x, 0) $, with corresponding
Hamiltonian ${\cal{H}}_{\rm{m}}^{\prime}  =   [{\mathbf{p}}  +  ({g \mu _{\rm{B}}}/{c}) {\mathbf{A}}_{\rm{m}}^{\prime}]^2/2m$.
Indeed, since $ U_g^{-1 } {\cal{H}}_{\rm{m}}^{\prime} U_g  = {\cal{H}}_{\rm{m}}$, with unitary transformation $U_g \equiv  {\rm{exp}}(i g \mu _{\rm{B}} {\cal{E}} xy/2 \hbar  c^2)$, the energy spectrum is the same for both gauges. We use now this gauge and drop the prime. 
The Landau gauge has the advantage of being constant in $y$ direction and thus we can perform a Fourier transformation of ${\cal{H}}_{\rm{m}}$ [Eq. (\ref{magnon_hopping})] in the variable $y$ and introduce the momentum $k_y$ such that ${\cal{H}}_{\rm{m}} =\sum_{k_y} H_{k_y}$, with
\begin{align}
H_{k_y}&=-t_x \sum_{n} (a_{k_y,n+1}^\dagger a_{k_y,n} + {\rm{H.c.}} ) \label{TightBinding2}   \\
            &\hspace{55pt}- 2 t_y  \sum_n  [ \cos (k_y a_y+\theta_n)] a_{k_y,n}^\dagger a_{k_y,n},         \nonumber 
\end{align}
where $a_{k_y,n}$ annihilates a magnon with momentum $k_y$ in $y$ direction at site $n=x/a_x$ (along $x$ direction), and where $a_{x(y)}$ is the lattice constant in $x(y)$ direction.
The phase accumulated by the magnon as it hops in  $y$ direction by  one lattice constant $a_y$ is given by  $\theta _{n} 
%=(g \mu _{\rm{B}}/\hbar c^2) \mathcal{E} n a_x a_y \equiv
= n \theta _0 $, where $\theta_0\equiv (g \mu _{\rm{B}}/\hbar c^2) \mathcal{E} a_x a_y$. For $t_x=t_y$, this is the standard Hamiltonian describing the integer QHE on a lattice in the presence of a flux. 
Note that for the spectrum the quantum statistics does not matter and the same spectrum is obtained for bosons and fermions. 

Next, we periodically extend ${\mathbf{A}}_{\rm{m}}^{\prime} = ({\mathcal{E}}/c)(0, x, 0) $  in $x$ direction, i.e., ${\mathbf{A}}_{\rm{m}}^{\prime} \to {\mathbf{A}}_{\rm{m}q}^{\prime} = ({\mathcal{E}} R_q/c) (0, \{x/R_q \}, 0) $, where  ${ R}_q$ is the period and $\{.\}$ means fractional part smaller than one. In the tight-binding model, it is convenient to choose $R_q= q a_x$ with $q$ being integer, which results in $\theta_n=\theta_0 q \{ n/q\}$. 
We want now to study the spectrum of the periodically extended $H_{k_y}$ (on a lattice strip oriented along $y$ direction and of finite width in $ x$ direction) and see how it depends on the period $R_q$. We obtain the spectrum by exact numerical diagonalization of the Hamiltonian~(\ref{TightBinding2}) and show the results in Fig. \ref{fig:Iso_BandSum} (see also Appendix \ref{sec:aniso}).
For definiteness, we will focus on the spectrum around the lowest Landau level. 
%%%%%%%%%%%%%%%%%%%%%%%%%%%%%%%%%%%%%%%%%%%

For large values of $q$ such that $ l_{\rm{{\mathcal{E}}}} \ll qa_x$ we see from Fig. \ref{fig:Iso_BandSum}(a) that
the bulk gap (between the first and second Landau level) is not flat, but still there is one chiral edge state propagating along $y$ direction at each edge of the strip. 
As  $q$ gets smaller, the bulk spectrum is modified, see Figs. \ref{fig:Iso_BandSum} (b)-(d). 
For even smaller values of $q$ such that $ l_{\rm{{\mathcal{E}}}} > qa_x$ the modification is very drastic in the sense that the bulk gap is closed for a fixed energy \cite{WeylMagnon} and the system becomes gapless; see Figs. \ref{fig:Iso_BandSum} (c) and (d).
However, quite remarkably, for each given value of $k_y$, there is still a gap in the spectrum, and, moreover, the chiral edge states still exist. Thus, if disorder is weak this edge modes will not couple to the bulk and the Hall conductance will still be dominated by these edge states, despite the fact that parts of the spectrum are gapless. 
This feature is similar to Weyl semimetals.
We refer to Appendix \ref{sec:aniso} for an anisotropic case $J_x  \not= J_y$.
Finally, we mention that we also tested numerically the periodic extension of the field gradient for the symmetric gauge and, as expected by gauge invariance, found the same behavior as for the Landau gauge.

%%%%%%%%%%%%%%%%%%%%%%%%%%%
\section{Hall conductances for magnons}
\label{sec:QHE}
%%%%%%%%%%%%%%%%%%%%%%%%%%%
\subsection{Magnetic Hall conductance}
\label{subsec:Bulk}
%%%%%%%%%%%%%%%%%%%%%%%%%%%
In this section we discuss the Hall transport properties of magnons. We begin with by introducing some relevant properties of the magnon spectrum applicable to a wide range of insulating magnets \cite{Mook2,Lifa}
where ${\cal{H}}_{\rm{m}}$ [Eqs. (\ref{magnon_hopping}) and (\ref{HamiltonianLL})] plays the role of an effective Hamiltonian for magnons (we suppress now the $z$ coordinate).
On the lattice, magnons are subject to a periodic lattice potential \cite{AMermin,Kohmoto,NiuBerry} $ U({\mathbf{r}}) = U({\mathbf{r}}+ {\mathbf{R}}) $ with Bravais lattice vector ${\mathbf{R}}=(a_x,a_y)$, and the total Hamiltonian is given by
$ {\cal{H}}({\mathbf{r}})   =   {\cal{H}}_{\rm{m}} ({\mathbf{r}})  + U({\mathbf{r}})  + g \mu _{\rm{B}} B_0$.
Following Refs. [\onlinecite{Kohmoto,NiuBerry}] we  introduce   the Bloch Hamiltonian with  Bloch wavevector ${\mathbf{k}}=(k_x, k_y) $,
$   {\cal{H}}_{\mathbf{k}} \equiv    {\rm{e}}^{- i {\mathbf{k}}\cdot{\mathbf{r}} } {\cal{H}}  {\rm{e}}^{ i {\mathbf{k}}\cdot{\mathbf{r}} }  = [- i \hbar  {\mathbf{\nabla}} +  \hbar  {\mathbf{k}}  + g \mu _{\rm{B}}{\mathbf{A}}_{\rm{m}}({\mathbf{r}})/c]^2/2m + U({\mathbf{r}})  + g \mu _{\rm{B}} B_0$.
The eigenfunction of the Schr\"odinger equation ${\cal{H}}_{{\mathbf{k}}} u_{n {\mathbf{k}}}({\mathbf{r}}) = E_{n {\mathbf{k}}} u_{n {\mathbf{k}}}({\mathbf{r}}) $
is given by \cite{KevinHallEffect,Kohmoto,NiuBerry} the magnonic Bloch wave function 
$ u_{n {\mathbf{k}}}({\mathbf{r}})  \equiv   {\rm{e}}^{- i {\mathbf{k}}\cdot{\mathbf{r}} }  \psi _{n {\mathbf{k}}}   $,
where ${\cal{H}} \psi _{n {\mathbf{k}}} = E_{n {\mathbf{k}}}  \psi _{n {\mathbf{k}}}  $
and $ E_{n {\mathbf{k}}}  \not=  E_{l {\mathbf{k}}}   $ when  $  n  \not= l  $; the eigenvalue  $E_{n {\mathbf{k}}} $ depends \cite{NiuBerry,Kohmoto} on ${\mathbf{k}}$ continuously.

We next focus on magnon Hall conductances.
As in Refs. [\onlinecite{VolovikQHE,Haldane2,Fujimoto,magnon2}], a gradient along, say, $x$ direction, of a magnetic field $B$ perpendicular to the $xy$ plane (Fig. \ref{fig:MagnonicSystem}) acts \cite{magnonWF,SilsbeeMagnetization,Basso,Basso2,Basso3} as a driving force for magnons like an electric field for charged particles.
Using this correspondence, we evaluate the magnon Hall conductance in the clean bulk limit generated by a weak constant magnetic field gradient $ |\partial _x B |  \ll  \hbar \omega _c/g\mu _{\rm{B}}a$, assuming that the system is surrounded by a large bulk magnet which acts as a reservoir \cite{magnon2} for magnons, providing effective chemical potentials for the magnons
\footnote{Therefore it can be assumed that magnons flow along a direction, but the lost magnon is compensated by the reservoir and the Bose-distribution function of magnons remains the equilibrium one; even without such a setup, the Bose-function qualitatively remains the same\cite{SilsbeeMagnetization,Basso,Basso2,Basso3} due to the energy gap in the magnon spectrum induced by $B$.}
In the bulk,  the anomalous velocity \cite{NiuBerry,Matsumoto,Matsumoto2,Fujimoto} associated with a confining potential is zero, and the calculation procedure indeed becomes in parallel with Ref. [\onlinecite{Kohmoto}], with the differences that we treat the magnons in the Landau quantization.

The Hall conductance $G ^{y x}$ of bulk magnons is defined by $ \langle  j_y \rangle = - G ^{y x} \partial _x B $, where  $ j_y  =  g\mu _{\rm{B}}v_y/L^2 $ is the magnon current density operator along the $y$-axis, $L^2$ being the area of the system, and the magnon velocity operator \cite{NiuBerry,Kohmoto,TKNN} $v_{y} =  \partial {\cal{H}}_{\mathbf{k}}/\partial \hbar  k_{y}$.
Focusing on the linear response regime, the magnon Hall transport is described by the transverse Kubo formula \cite{NiuBerry,Kohmoto,TKNN}  (see Appendix \ref{sec:BulkMagnonConductivity} for details),
\begin{eqnarray}
G ^{y x} &=&  \frac{(g \mu _{\rm{B}})^2}{h}  \sum_n 
  \int_{{\rm{BZ}}}  \frac{d^2 k}{2\pi} n_{\rm{B}} (E_{n {\mathbf{k}}})  {\Omega} _{n, z}(\mathbf{k}),  
\label{QHE} 
\end{eqnarray}
where $ n_{\rm{B}} (E_{n {\mathbf{k}}})=({\rm{e}}^{\beta E_{n {\mathbf{k}}}}-1)^{-1}$ with $\beta \equiv (k_{\rm{B}}T)^{-1} $ is the Bose-distribution function, $\rm{BZ}$ denotes the corresponding Brillouin zone  analogous to the magnetic \cite{NiuBerry,Kohmoto,Zak} BZ for electronic systems in the presence of a magnetic flux, and ${{\mathbf{\Omega}}}_{n}(\mathbf{k})$ is the Berry curvature (see Appendix \ref{sec:BulkMagnonConductivity} for details).
We note that the time reversal symmetry of the system is broken by the magnetic field $B$  perpendicular to the $xy$ plane and by the finite magnetization. In such systems with broken time reversal symmetry, the Berry curvature generally becomes \cite{KevinHallEffect,Matsumoto,Matsumoto2,NiuBerry} non-zero.
Thus, the magnon Hall conductance is generally characterized by the product of the Berry curvature and the Bose-distribution function as shown in Eq. (\ref{QHE}).

Finally, we provide the condition for QHEs characterized by a topological invariant, i.e., by a Chern number \cite{Kohmoto,TKNN} associated with the Berry curvature.
It may be assumed that the energy level spacing [see Eq. (\ref{LL})] is characterized by 
$ \mid   E_{n {\mathbf{k}}}  -    E_{{n-1} {\mathbf{k}}}  \mid   \sim   \hbar \omega _c  $.
%%%%%%%%%%%%%%%%%%%%%%%%%%%%%%%%%%%%%%%%%%%%%%%
At low temperature $k_{{\rm{B}}} T  \ll  \hbar \omega _c$, only the lowest mode $ n=0$ becomes relevant in Eq. (\ref{QHE}).
We can then consider the case \cite{KevinHallEffect,RSdisorder},
$ \mid  {\rm{max}}   \{ E_{0 {\mathbf{k}}}:  {{\mathbf{k}}\in ({\rm{BZ}})} \}  
-   {\rm{min}}   \{ E_{0 {\mathbf{k}}}:  {{\mathbf{k}}\in ({\rm{BZ}})} \} \mid    \ll    k_{{\rm{B}}} T $ 
with $ \partial _k  E_{0 {\mathbf{k}}} \not=0  $, where the band width is much smaller than $  k_{{\rm{B}}} T$ and the lowest energy band can be regarded as being {{almost}} flat in the Bloch wavevector-space.
For such an {{almost flat band}}, Eq. (\ref{QHE}) becomes
\begin{eqnarray}
G ^{y x}    \approx    \frac{(g \mu _{\rm{B}})^2}{h} n_{\rm{B}} (E_0^{\ast })  \cdot   \nu _0,    
\label{QHE_zeromode} 
\end{eqnarray}
with $\nu _0  \equiv  \int_{{\rm{BZ}}}  ({d^2 k}/{2\pi}) {\Omega} _{0, z}(\mathbf{k}) \in  {\mathbb{Z}}$,  
where $E_0^{\ast }$ represents the typical~\footnote{
$ E_0^{\ast }$ can be represented by any value
%$ E_0^{\ast } = ^{\exists}
$E_{0 {\mathbf{k}}}$ in the the almost flat band interval
$[{\rm{min}_{\mathbf{k}}}  \{ E_{0 {\mathbf{k}}}\}, {\rm{max}_{\mathbf{k}}}  \{ E_{0 {\mathbf{k}}}\} ]  $,
since the value of $n_{\rm{B}} (E_0^{\ast }) $ approximately remains constant in this interval.
}
energy value for the almost flat band and $\nu _0 $ is the Chern number \cite{Mook,Mook2,Mook3,Lifa}. 
Due to the single-valuedness of the wave function, it takes on integer values, \cite{Kohmoto} $\nu _0 \in  {\mathbb{Z}} $.
%%%%%%%%%%%%%%%%%%%%%%%%%%%%%%%%%%%%%%%%%%%%%%%%%%%%%%%%%
Thus, in the almost flat band, 
\begin{eqnarray}
({\rm{Band  \   width}}) \ll   k_{{\rm{B}}} T  \ll  \hbar \omega _c   \   \    {\rm{with}}   \ \partial _k  E_{0 {\mathbf{k}}} \not=0,
\label{AlmostFlat} 
\end{eqnarray}
the conductance at low temperature becomes characterized by the Chern number $\nu _0$.
Eq. (\ref{QHE_zeromode}) indicates that at such low temperatures the magnetic Hall conductance of magnons in the clean bulk could be regarded as being quantized in units of  $ [(g \mu _{\rm{B}})^2/{h}]  n_{\rm{B}} (E_0^{\ast })$.
This is analogous to the (integer) QHE \cite{TKNN,Kohmoto} of charged particles quantized  in units of $e^2/{h}$ where the Fermi-distribution function is replaced by the Heaviside step function at zero temperature.
%%%%%%%%%
We emphasize that  the Hall conductance for magnons, however, depends on temperature and on the typical energy value for the almost flat band via the Bose-distribution function $n_{\rm{B}} (E_0^{\ast })$.
This arises from the intrinsic bosonic nature of magnons and the fact that $n_{\rm{B}} (E_0^{\ast })$ behaves fundamentally different from the Fermi distribution at low temperatures
\footnote{See also Appendix \ref{sec:BulkMagnonConductivity} for the difference between the magnonic QHE and the one in a disordered chiral spin liquid \cite{Haldane2} where the Bloch wavevector is no longer a good quantum number due to disorder \cite{NiuPhaseTwist} effects ({\it{e.g.}}, by impurities).};
due to the Bose function in Eq. ({\ref{QHE_zeromode}}), the  Hall conductance of noninteracting magnon vanishes at zero temperature. 
This result is fully consistent with the general conclusion given in Ref. [\onlinecite{TaoMaki}] that there cannot be any transport signature of the  QHE for non-interacting bosons.
In other words,  the magnonic QHE manifests itself with a finite Hall conductance only at finite temperatures, as described by Eq. (11).

%
%
%This exactly agrees with the argument by Tao and Maki \cite{TaoMaki}. 
%Thus at finite non-zero temperature, magnonic QHE defined via its quantized Hall conductance is realized in the almost flat band [Eq. (\ref{AlmostFlat})] where we assume explicitly finite non-zero temperature.
%%%%%%%%%%%%%%%%%%%%%%%%%%%%%%%%%%%%%%%%%%%%%%%%%%%%%%%
We note a sum rule \cite{RShindou,Mook2} for Chern number and when the (lowest) band becomes {\textit{completely}} flat  $ \partial _k  E_{n {\mathbf{k}}} =0  $,
the Hall conductance can become zero since the Berry curvature itself \cite{Matsumoto,Matsumoto2} vanishes.
Such  almost flat bands [Eq. (\ref{AlmostFlat})], for instance, are  realized \cite{KevinHallEffect} in a skyrmion lattice induced by the Dzyaloshinskii-Moriya \cite{DM,DM2,DM3} (DM) interaction which provides \cite{katsura2,Mook2,Lifa}  an effective AC phase. \footnote{Also in Refs. [\onlinecite{Mook2,Lifa}], the AC phase of the form of Eq. (\ref{eqn:SpinHamiltonian}) is induced \cite{katsura2} by the DM interaction.}
In particular, the DM interaction produces \cite{KevinHallEffect} a textured equilibrium magnetization that acts intrinsically as a vector potential analogous to ${\mathbf{A}}_{\rm{m}}$ [Eqs. (\ref{eqn:SpinHamiltonian}) and (\ref{magnon_gauge})]. In the skyrmion lattice, the low-energy magnetic excitations  are magnons and the Hamiltonian indeed reduces \cite{KevinHallEffect} to the same form of Eq. (\ref{HamiltonianLL}) where the analog of the Landau gauge, 
${\mathbf{A}}_{\rm{m}}^{\rm{sky}} ({\mathbf{r}}) = -  {\cal{B}}_0  y  {\mathbf{e_x}}$, is provided by the average fictitious magnetic field ${\cal{B}}_0 = 2 \pi/\sqrt{3}{\cal{R}}^2 $ for a skyrmion of radius ${\cal{R}}$ on top of a periodic contribution with zero average which induces an almost flat band in the magnon spectrum \cite{KevinHallEffect}.

%%%%%%%%%%%%%%%%%%%%%%%%%%%
\subsection{Thermal Hall conductance}
\label{subsec:edge}
%%%%%%%%%%%%%%%%%%%%%%%%%%%
We apply the above results to the thermal Hall conductance \cite{Matsumoto,Matsumoto2}.
Focusing on systems with boundaries, we consider a magnetic insulator in the absence of any externally applied magnetic field gradients (i.e., ${\mathbf{\nabla}}  B =0 $), 
while a thermal gradient along $x$ direction $ \partial _x T \not=0$ is sustained by contacts to thermal baths of different temperatures, see Fig. \ref{fig:MagnonicSystem}.
We work under the assumption  that the spin along the $z$ direction is a good quantum number.  Within linear response theory, magnon and heat current densities, $ {\mathbf{j}}$ and $ {\mathbf{j}}^Q $, respectively, are then generally characterized by the following $4\times 4$ Onsager matrix, 
\begin{eqnarray}
%%%%%%%%%%%%%%%%%%%%%%%%%%%%%
\begin{pmatrix}
\langle j_x \rangle  \\   \langle j_y \rangle  \\  \langle  j_x^{Q}   \rangle  \\  \langle  j_y^{Q}   \rangle
\end{pmatrix}
%%%%%%%%%%%%%%%%%%%%%%%%%%%%%
=
%%%%%%%%%%%%%%%%%%%%%%%%%%%%%
\begin{pmatrix}
L_{11}^{xx} & L_{11}^{xy}  & L_{12}^{xx}  & L_{12}^{xy}    \\    
L_{11}^{yx} & L_{11}^{yy}  & L_{12}^{yx}  & L_{12}^{yy}    \\    
L_{21}^{xx} & L_{21}^{xy}  & L_{22}^{xx}  & L_{22}^{xy}    \\    
L_{21}^{yx} & L_{21}^{yy}  & L_{22}^{yx}  & L_{22}^{yy}    
\end{pmatrix}
%%%%%%%%%%%%%%%%%%%%%%%%%%%%%
%%%%%%%%%%%%%%%%%%%%%%%%%%%%%
\begin{pmatrix}
-   \partial _x  B  \\  -   \partial _y  B  \\ -  \partial _x  T/T  \\ -  \partial _y  T/T
\end{pmatrix}.
%%%%%%%%%%%%%%%%%%%%%%%%%%%%%
\label{eqn:4by4}
\end{eqnarray}
Here, the temperature and field gradients are all in general non-zero,  containing contributions from external and internal sources,
where the latter are generated in the stationary state by the Hall effect itself.
%%%%%%%%%%%%%%%%%%%%%%%%%%%
Focusing on the quantum Hall regime $| L_{i j}^{xx}  |  \ll  | L_{i j}^{yx}  |  $ ($i, j = 1, 2$),
the Hall current densities are given by the Hall coefficients $L_{i j}^{yx}$ in leading order.  If the bands are fully gapped, then $L_{i j}^{\mu \mu} =0$ exactly. For bands like in Fig.~\ref{fig:Iso_BandSum}~(c-d) this is no longer the case, but as long as the spectral part around the edge states is gapped, the contribution to transport
 from the extended continuum states will lead to small corrections. \footnote{The longitudinal resistance $R^{\mu\mu}$ coming from ungapped continuum states (without disorder) is much smaller than the Hall resistance $R^{xy}\sim h/(g\mu_{\rm{B}})^2$ of a few edge modes, and thus $G^{xy}=R^{xy}/[(R^{xx})^2+(R^{xy})^2]=(1/R^{xy})[1 + {\cal{O}}\big((R^{xx}/R^{xy})^2\big)]$. Similarly, $G^{xx}=R^{xx}/[(R^{xx})^2+(R^{xy})^2]\approx R^{xx}/(R^{xy})^2 \ll G^{xy}$ for $R^{xy}\gg R^{xx}$.
  This means that for the Hall conductance the longitudinal bulk conductance contributions can be neglected.}
  %%%%%%%%%%%%%%%%%%%
 If the band is almost flat around the region of the spectrum containing the edge modes, $L_{i j}^{yx}$ are well approximated by the Chern number and given by Eq. (\ref{QHE_zeromode}).
%%%%%%%%%%%%%%%%%%%%%%%%%%%%%%%%%
Applying then the condition of flat bands Eq. (\ref{AlmostFlat}) also to the thermal transport coefficients obtained in Ref. [\onlinecite{Matsumoto}], we readily obtain
\begin{eqnarray}
L_{i j}^{yx} =   (k_{\rm{B}}T)^{\eta} (g \mu _{\rm{B}})^{2-{\eta}}  {\cal{C}}_{\eta}  \big(n_{\rm{B}} (E_0^{\ast })\big) \cdot  \nu _0/h,
 \label{eqn:HallLij}
\end{eqnarray}
where
$  L_{1 1}^{yx} = G ^{yx} $ [Eq. (\ref{QHE_zeromode})],
$ \eta =  i + j -2$, 
${\cal{C}}_0  \big(n_{\rm{B}} (E_0^{\ast })\big)  =   n_{\rm{B}} (E_0^{\ast }) $,
${\cal{C}}_1  \big(n_{\rm{B}} (E_0^{\ast })\big)  =  [1+ n_{\rm{B}} (E_0^{\ast })] {\rm{log}} [1+ n_{\rm{B}} (E_0^{\ast })] 
-  n_{\rm{B}} (E_0^{\ast }) {\rm{log}} [n_{\rm{B}} (E_0^{\ast })] $, and
${\cal{C}}_2  \big(n_{\rm{B}} (E_0^{\ast })\big)  =   [1+ n_{\rm{B}} (E_0^{\ast })]  \big({\rm{log}} [1+ 1/n_{\rm{B}} (E_0^{\ast })]\big)^2 - \big({\rm{log}} [n_{\rm{B}} (E_0^{\ast })]\big)^2 - 2 {\rm{Li}}_2 \big(- n_{\rm{B}} (E_0^{\ast })\big)$ with the polylogarithm function  ${\rm{Li}}_{s}(z) = \sum_{n=1}^{\infty} z^n/n^s$.

\begin{figure}[h]
\begin{center}
\includegraphics[width=8.8cm,clip]{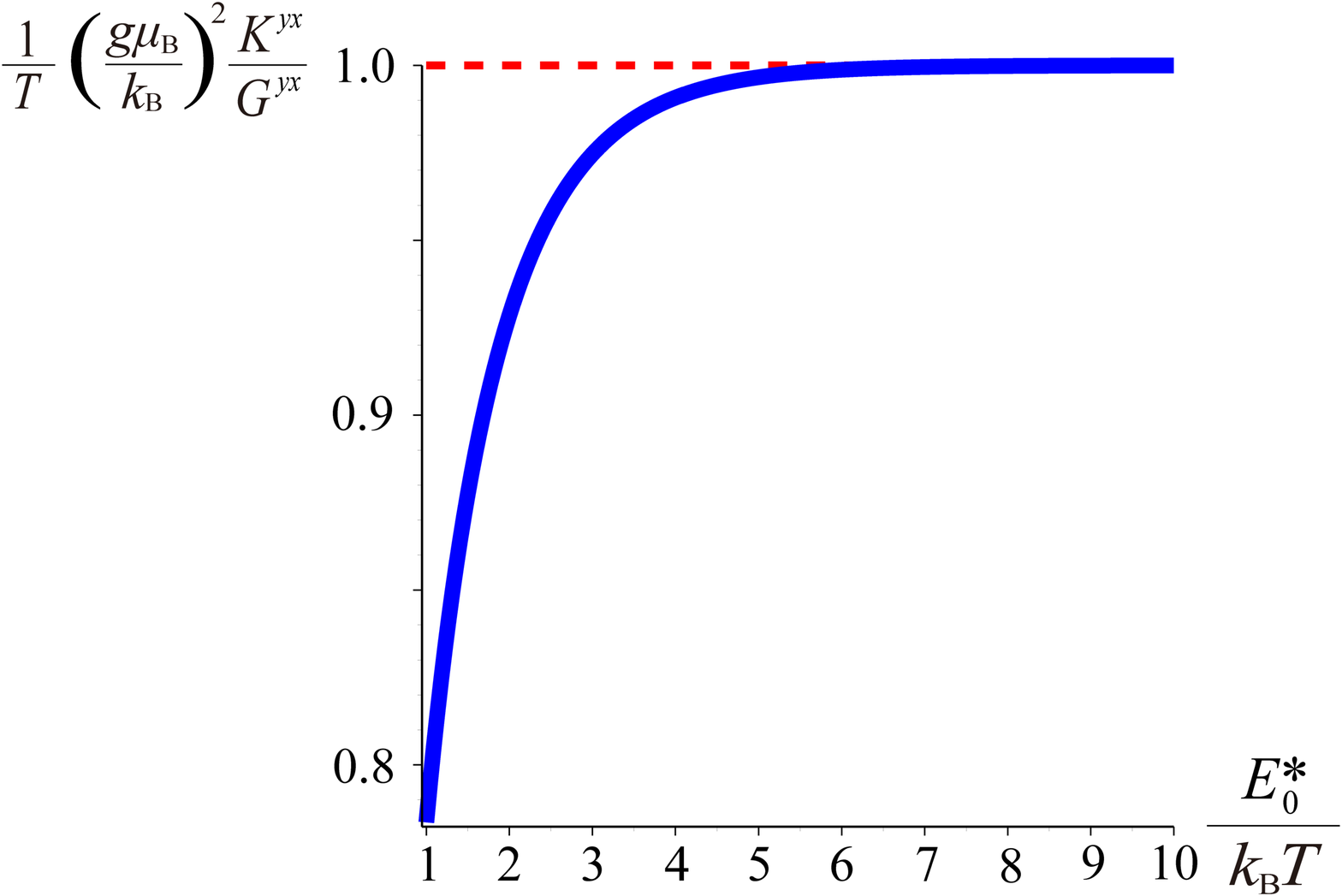}
\caption{(Color online)
Plot of the ratio $ (g \mu _{\rm{B}}/k_{\rm{B}})^2 (K^{yx}/G ^{yx}T)$ as function of $  E_0^{\ast }/k_{\rm{B}}T$.
At low temperatures $ E_0^{\ast }/k_{\rm{B}}T \geq 5 $, the ratio becomes constant and the magnonic WF law [Eq. (\ref{eqn:WFQHE})] is realized. 
See also Fig. \ref{fig:WFedgeQHE2}.
\label{fig:WFedgeQHE} }
\end{center}
\end{figure}

Focusing on the Hall transport along $y$ direction,  and assuming $  |L_{i j}^{xx} /  L_{i j}^{yx}|  \ll  1 $, the applied thermal gradient $ \partial _x  T$ induces a magnon Hall current $\langle j_y \rangle  = - L_{12}^{yx} \partial _x  T/T$, which leads to an accumulation of magnons at the boundaries and thereby builds up a non-uniform magnetization in the sample. This in turn generates a intrinsic magnetization gradient \cite{SilsbeeMagnetization,magnonWF, Basso,Basso2,Basso3} along both directions, and that along $x$ direction $\partial _x  B^{\ast }$ produces a magnon counter  Hall current along $y$ direction.
Then, the system reaches a new stationary  state such that in- and out-flowing magnon currents along y direction balance each other, and there is no total magnon current in this new quasi-equilibrium state, i.e., $\langle j_y \rangle = 0$. 
This is the case when
\begin{eqnarray}
   \partial _x  B^{\ast } = -  \frac{L_{12}^{yx}}{L_{11}^{yx}}   \frac{    \partial _x  T}{T}.
  \label{eqn:NewState}
\end{eqnarray}
The thermal Hall conductance $K^{yx}$,  defined by $\langle   j_y^Q   \rangle = - K^{yx} \partial _x  T$, is measured under this condition. 
This is in complete analogy to  thermal transport of electrons in metals \cite{AMermin}.
Thus, putting Eq. (\ref{eqn:NewState}) into Eq. (\ref{eqn:4by4}), the thermal Hall conductance $K^{yx}$ expressed in terms of Onsager coefficients  becomes
\begin{eqnarray}
    K^{yx} = \Big( L_{22}^{yx} - \frac{L_{21}^{yx} L_{12}^{yx}}{L_{11}^{yx}}   \Big)/T,
  \label{eqn:K_magnon}
\end{eqnarray}
where the off-diagonal elements arise from the magnon counter-current. 
From this we obtain the thermomagnetic ratio $ {K^{yx}}/{G ^{yx}} $, characterizing  magnon and heat Hall transport.
This ratio is plotted in Fig. \ref{fig:WFedgeQHE}.
At low temperatures (i.e., $E_0^{\ast }/k_{\rm{B}}T \geq 5 $), the ratio of the non-dissipative transverse transport coefficients becomes linear in temperature,
\begin{eqnarray}
 \frac{K^{yx}}{G^{yx}} = \Big(\frac{k_{\rm{B}}}{g \mu _{\rm{B}}}\Big)^2 T \Big[\frac{{\cal{C}}_2}{{\cal{C}}_0} - \Big(\frac{{\cal{C}}_1}{{\cal{C}}_0}\Big)^2 \Big]     
\stackrel{\rightarrow }{=}  \Big(\frac{k_{\rm{B}}}{g \mu _{\rm{B}}}\Big)^2 T
 \label{eqn:WFQHE}
\end{eqnarray}
with a universal proportionality constant 
\begin{eqnarray}
{\cal{L}}= \Big(\frac{k_{\rm{B}}}{g \mu _{\rm{B}}}\Big)^2,
 \label{eqn:Lorenz}
\end{eqnarray}
which we refer to as magnetic Lorenz number~\cite{magnonWF}.
Instead of the charge $e$, ${\cal{L}}$ is characterized by $g \mu _{\rm{B}}$ and it is independent of any geometry and material parameters \footnote{We note that adding a confining potential to Eq. (\ref{eqn:4by4}) as in Refs. [\onlinecite{Matsumoto,Matsumoto2}], and thus considering the edge currents, the magnonic WF law given in Eq. (\ref{eqn:WFQHE}) still holds.} except for the $g$-factor.
Thus, at low temperatures, the ratio of the magnonic quantum Hall conductances satisfies the WF law.
%%%%%%%%%%%%%%%%%%%%%%%
Interestingly, the law holds in the same way for topologically non-trivial QHE systems in two dimensions as well as for  three-dimensional ferromagnetic insulating junctions 
that are topologically trivial \cite{magnonWF}. 
%magnon transport in two-dimensional quantum Hall systems as for topologically trivial one in three-dimensional ferromagnetic insulating junctions \cite{magnonWF} that is essentially classical phenomenon.
This is another manifestation of the universality of the  WF law.
% does not depend even on the dimensionality of the systems.
%%%%%%%%%%%%%%%%%%%%%%%
This is one of the main results of this work.
%%%%%%%%%%%%%%%%
We note that magnon and heat currents are generally characterized by the $4 \times  4$ Onsager matrix Eq. (\ref{eqn:4by4}), 
and without the quantized Hall conductance given in Eq. (\ref{AlmostFlat}),
the expression for the thermal Hall conductance  in Eq. (\ref{eqn:K_magnon}) drastically changes due to the longitudinal transport coefficients $L_{i j}^{\mu \mu} $ 
and as a consequence the WF law cannot be recovered in such a classical Hall regime.

\begin{figure}[h]
\begin{center}
\includegraphics[width=8.2cm,clip]{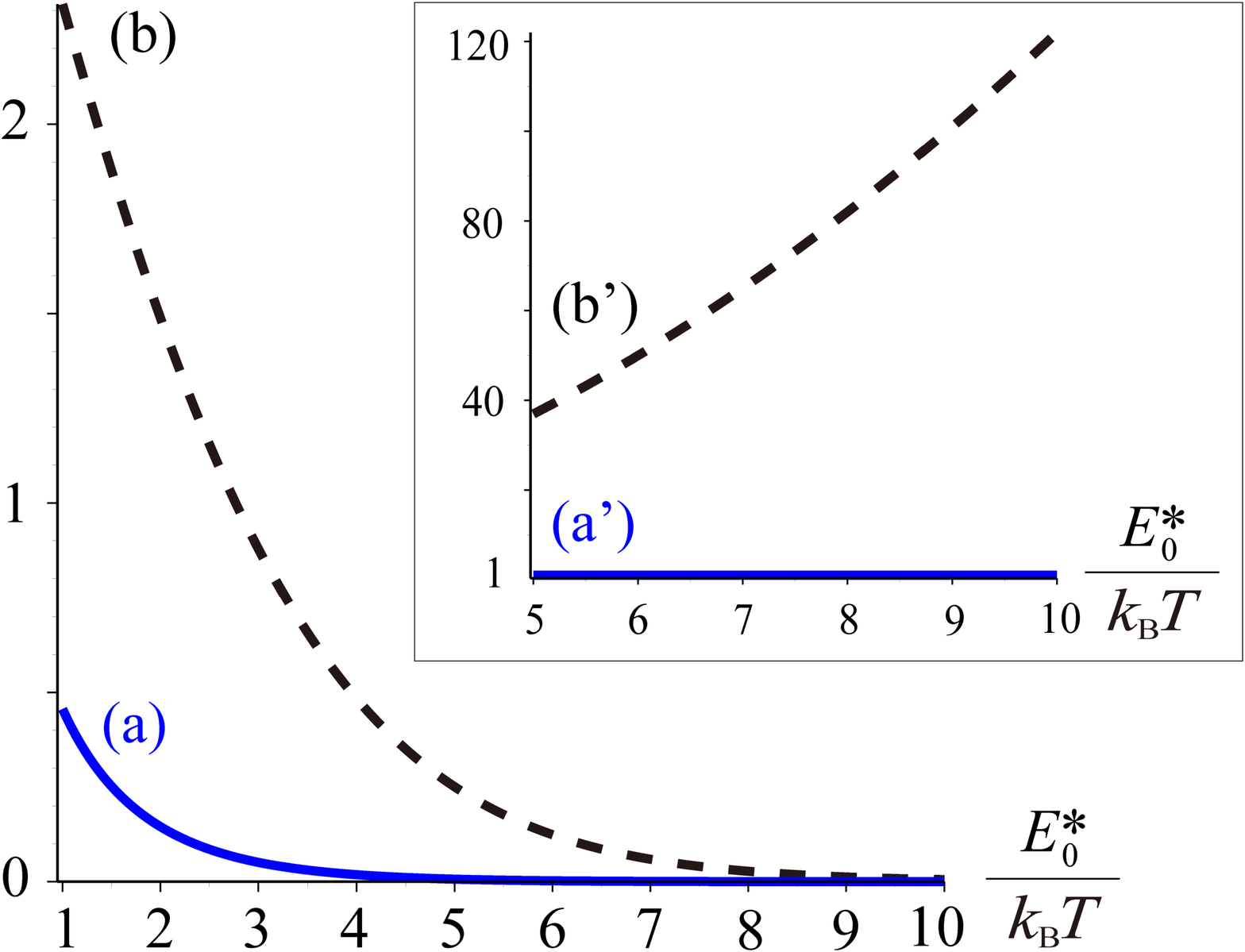}
\caption{(Color online)
Plots of 
(a) $  h K^{yx}/k_{\rm{B}}^2 T $ and
(b) $  h \kappa ^{yx}/k_{\rm{B}}^2 T $ as function of $  E_0^{\ast }/k_{\rm{B}}T$ with assuming $\nu _0 = 1$.
The deviation remains substantial even at  low temperatures.
Inset: Plots of the ratio
(a') $  (g \mu _{\rm{B}}/k_{\rm{B}})^2 (K^{yx}/G ^{yx}T)$ and 
(b') $  (g \mu _{\rm{B}}/k_{\rm{B}})^2 (\kappa^{yx}/G ^{yx}T)$.
In contrast to (a'),  the ratio (b') does not reduce to a constant even at low temperatures.
\label{fig:WFedgeQHE2} }
\end{center}
\end{figure}

Remarkably, the WF law \footnote{As to the electronic WF law for diagonal ($L_{i j}^{xx}$) and non-diagonal ($L_{i j}^{yx}$) transport coefficients in impurity-disordered quantum Hall systems, see Refs. [\onlinecite{GrunwaldHajdu,KaravolasTriberis}].}
holds in the same way for magnons, which are bosons, as for electrons \cite{WFgermany} which are fermions. 
However, there is a crucial difference in the thermal conductances. 
For electron transport \cite{AMermin,LandauWF} the thermal conductance $K$ may be approximately identified with the diagonal Onsager coefficient $L_{22}/T \equiv  \kappa$  since electrons have a sharp Fermi surface at the Fermi energy  $\epsilon _{\rm{F}}$ resulting in a strong suppression of off-diagonal contributions,
$ K - \kappa  \propto  {\cal{O}} \big( (k_{\rm{B}}T/\epsilon _{\rm{F}})^2  \big) \ll 1$ even at room temperature where   $\epsilon _{\rm{F}} \gg  k_{\rm{B}}T$ still holds 
for typical metals. The same applies generally for fermionic excitations with sharp Fermi surface.
%%%%%%%%%%%%%%%
However, it is obvious that such an approximation is not valid for magnons, which are characterized by the Bose distribution, and indeed breaks down, see Fig. \ref{fig:WFedgeQHE2}. 
There we plot $ \kappa ^{yx} \equiv  L_{22}^{yx}/T =  (k_{\rm{B}}^2 T/h) {\cal{C}}_2 \cdot  \nu _0$ as function of inverse temperature,
which shows that even at low temperatures there remains a sizable difference from the properly defined thermal Hall conductance $K^{yx} $ [Fig. \ref{fig:WFedgeQHE2} (a)]
since the off-diagonal coefficients in Eq. (\ref{eqn:K_magnon}) are as large as the diagonal ones.
Thus, the ratio $ \kappa ^{yx}/G ^{yx}$  does not obey a WF law~\footnote{We have confirmed this also by analytic calculation.}, see Fig. \ref{fig:WFedgeQHE2} (b').

%Finally, we wish to emphasize that our derivation of the WF law is only valid under the assumption that the $z$ component of the spin is a good quantum number. In particular, this excludes
%spin-lattice relaxation processes and magnon-magnon scatterings. For this we need to assume low enough temperatures, typically a few Kelvins and below,
%%At sufficiently low temperatures, effects of magnon-magnon interactions and phonons 
%where such processes become \cite{magnonWF,adachiphonon} negligibly small.

%%%%%%%%%%%%%%%%%%%%%%%%%%%
\section{Estimates for experiments}
\label{sec:experiments}
%%%%%%%%%%%%%%%%%%%%%%%%%%%
The magnonic Hall currents could be experimentally observed by Brillouin light scattering spectroscopy \cite{demokritovReport,demokritov,ultrahot,SergaBEC,MagnonPhonon,onose}.
For an estimate, we assume the following parameter values, 
${\mathcal{E}} = 1$V/nm$^2$, $J=30 $meV, $S=10$, $g=2$, and $a = 1$\AA ($10$\AA).
This provides the Landau gap $ \hbar \omega _c  =  0.01 \mu $eV ($1 \mu$eV) and $ l_{\rm{{\mathcal{E}}}}= 0.7 \mu $m [Eqs. (\ref{omega_c}) and (\ref{l_E})].
Therefore, the magnonic QHE could be observed at $T \lesssim 0.1 $mK ($10$mK) and the Chern number can be changed as function of the electric field gradient.
Note that at such low temperatures effects of magnon-magnon interactions and phonons become negligible \cite{magnonWF,adachiphonon}.
%%%%%%%%%%%%%%%%%%
These are rather low temperatures. A more realistic situation is obtained for skyrmion lattices \cite{KevinHallEffect}.
As already mentioned above, in these systems the DM interaction produces \cite{Mook2,Lifa,katsura2} intrinsically a vector potential analogous to ${\mathbf{A}}_{\rm{m}}$ [see Eq. (\ref{HamiltonianLL}) and also refer to Sec. \ref{subsec:Bulk}]; 
further, the low-energy magnetic excitations in the skyrmion lattice are magnons and the Hamiltonian indeed reduces to the same form of Eq. (\ref{HamiltonianLL}),
giving an almost flat band [Eq. (\ref{AlmostFlat})]. 
%%%%%%%%%%%%%%%%%%
For an estimate, we assume the following experimental parameter values \cite{SkyrmionReviewNagaosa,SkyrmionExpTokura,SkyrmionTheory},
$J=80 $meV, $a = 10$\AA, the radius of a skyrmion ${\cal{R}}=15 $nm, and the DM interaction $D=0.7$meV,
which provides a Landau gap of $ 2.5$meV.
Therefore, the magnonic QHE could be observed at  $T \lesssim 25 $K and the Chern number could be varied \cite{Mook2,Lifa} as a function of the DM interaction. 
The temperature, however, should be low enough to make spin-phonon and magnon-magnon contributions negligible \cite{magnonWF,adachiphonon}.
%%%%%%

We note that plateaus in the Hall current versus electric field gradient could be realized by injecting magnons into the system at an energy $E_{\rm{inj}}$ 
inside the gap between subsequent 
Landau levels $E_{n}^{\ast }$.  These magnons will then populate the chiral edge states on each edge and propagate along the edges, giving rise to a Hall current quantized in units
of $\nu _0(g\mu_B)^2/h $, with the Chern number $\nu _0$ corresponding to the number of edge states. For instance, $\nu _0 =1$ when $E_0^{\ast } < E_{\rm{inj}} < E_1^{\ast }$, while $\nu _0 =0$ when  $ E_{\rm{inj}} < E_0^{\ast }$. Bulk-edge correspondence ensures that as long as the chiral edge states exist, the quantization is robust against disorder effects \cite{Haldane2,RSdisorder}.

%As to plateaus, for instance, injecting a magnon with energy $E_{\rm{inj}}$, it may be realized as function of electric field gradient that characterizes the typical energy for each Landau level $E_{0(1)}^{\ast }$; the energy band spectrum (Fig. \ref{fig:Iso_BandSum}) indicates that the injected magnon becomes the chiral edge mode and contributes to the Chern integer $\nu _0 =1$ when $E_0^{\ast } < E_{\rm{inj}} < E_1^{\ast }$, while $\nu _0 =0$ when  $ E_{\rm{inj}} < E_0^{\ast }$.
%Bulk-edge correspondence ensures that as far as the chiral edge magnon state exists, the quantization is robust against disorder effects \cite{Haldane2,RSdisorder}.
%%%%%%%%%%%%%%%%%
Given these estimates, we conclude that the  observation of magnonic QHEs and the WF law, while being challenging, 
seems within experimental reach \cite{demokritov,ultrahot,SergaBEC,spinwave,ACspinwave,demokritovReport,MagnonPhonon,SkyrmionReviewNagaosa,SkyrmionExpTokura,Snell_Exp,Exp_TopMagnon}.

%%%%%%%%%%%%%%%%% 
\section{Summary}
\label{sec:sum}
%%%%%%%%%%%%%%%%%
We have studied the Aharonov-Casher effect on topological magnon transport and proposed a magnonic quantum Hall effect in Landau quantization at finite temperature
for quadratic and linear (Dirac-like) dispersion relations of magnons.
Moving magnons in a skewed-harmonic electric potential, or alternatively in a skyrmion lattice induced by the Dzyaloshinskii-Moriya interaction, give rise to  Landau level quantization and the Hall conductances become characterized by the topological Chern number for almost flat bands.
We showed that the quantum Hall features remain largely intact even if the effective flux (generated by an electric saw-tooth gradient)  is periodic in space with a period smaller than the electric Hall length. 
%%%%%%%%%%%%%%
We found that for temperatures lower than the Landau gap, the quantized Hall conductances obey an analog of the Wiedemann-Franz law where the ratio of heat to magnon conductance is linear in temperature and is $\it{universal}$, i.e., independent of geometry and material parameters of the system.
%%%%%%%%%%%%%%
It is well-known that quantum-statistical properties of bosons and fermions are fundamentally different and quantum effects become dominant in particular in the low temperature regime.
However, appropriately defining the thermal conductance of magnons with taking into account magnon counter-currents induced by magnetization gradients,
we discovered that transport in quantum Hall system exhibits the same linear-in-$T$ behavior as fermions. 
This is another demonstration of  the universality of the Wiedemann-Franz law independent of  particle statistics.

\begin{acknowledgments}
We acknowledge support by the Swiss National Science Foundation and the NCCR QSIT.
One of the authors (K.N.) gratefully acknowledges support by the JSPS (Fellow No. 26-143).
We thank S. Nigg,  C. Schrade, R. Tiwari, S. Hoffman, and A. Zyuzin for helpful discussions.
\end{acknowledgments}

\appendix

%%%%%%%%%%%%%%%%%%%%%%%%%%% 
\section{Magnonic Landau level}
\label{sec:MagnonicLLcal}
%%%%%%%%%%%%%%%%%%%%%%%%%%%
In this Appendix, we provide some  details of the straightforward calculation for the Landau level of nonrelativistic-like magnons for completeness.
Using the analogy explained in the main text, the calculation becomes analogous to the one for electrons\cite{mahan,Ezawa}.
Introducing operators analogous to a covariant momentum $ {\mathbf{\Pi}} \equiv   {\mathbf{p}}  +  g \mu _{\rm{B}}{\mathbf{A}}_{\rm{m}} /c  $,
which satisfy $[\Pi_x, \Pi_y]=i \hbar ^2/ l_{\rm{{\mathcal{E}}}}^2  $, the Hamiltonian (\ref{HamiltonianLL}) can be rewritten as 
$ {\cal{H}}_{\rm{m}}  = (\Pi_x^2 + \Pi_y^2)/2m $.
Next, introducing  the operators $ a \equiv     l_{\rm{{\mathcal{E}}}} (\Pi_x + i \Pi_y)/\sqrt{2} \hbar  $ and $a^{\dagger }\equiv     l_{\rm{{\mathcal{E}}}} (\Pi_x - i \Pi_y)/\sqrt{2} \hbar $, which satisfy bosonic commutation relations, i.e., $ [a, a^{\dagger }] = 1$ and the rest commutes, the Hamiltonian becomes 
 $ {\cal{H}}_{\rm{m}} =  \hbar  \omega _c  (a^{\dagger } a + 1/2)$.
Indeed, introducing \cite{Ezawa} the guiding-center coordinate by $ X =  x  +  l_{\rm{{\mathcal{E}}}}^2 \Pi_y/\hbar   $ and $ Y = y  -  l_{\rm{{\mathcal{E}}}}^2 \Pi_x/\hbar  $, which satisfy $[X, Y]= - i  l_{\rm{{\mathcal{E}}}}^2$ with $ dX/dt = dY/dt=0 $,  the time-evolution of the relative coordinate
 ${\mathbf{R}}_{\rm{{\mathcal{E}}}} =({\cal{R}}_x, {\cal{R}}_y) \equiv  (-  l_{\rm{{\mathcal{E}}}}^2 \Pi_y/\hbar,  l_{\rm{{\mathcal{E}}}}^2 \Pi_x/\hbar)$ becomes 
$  d({\cal{R}}_x + i {\cal{R}}_y) /dt = - i \omega _c  ({\cal{R}}_x + i {\cal{R}}_y)$.
Thus, magnons perform cyclotron motion and form Landau levels in the presence of electric field gradients.

\begin{figure}[t]
\begin{center}
\includegraphics[width=8.5cm,clip]{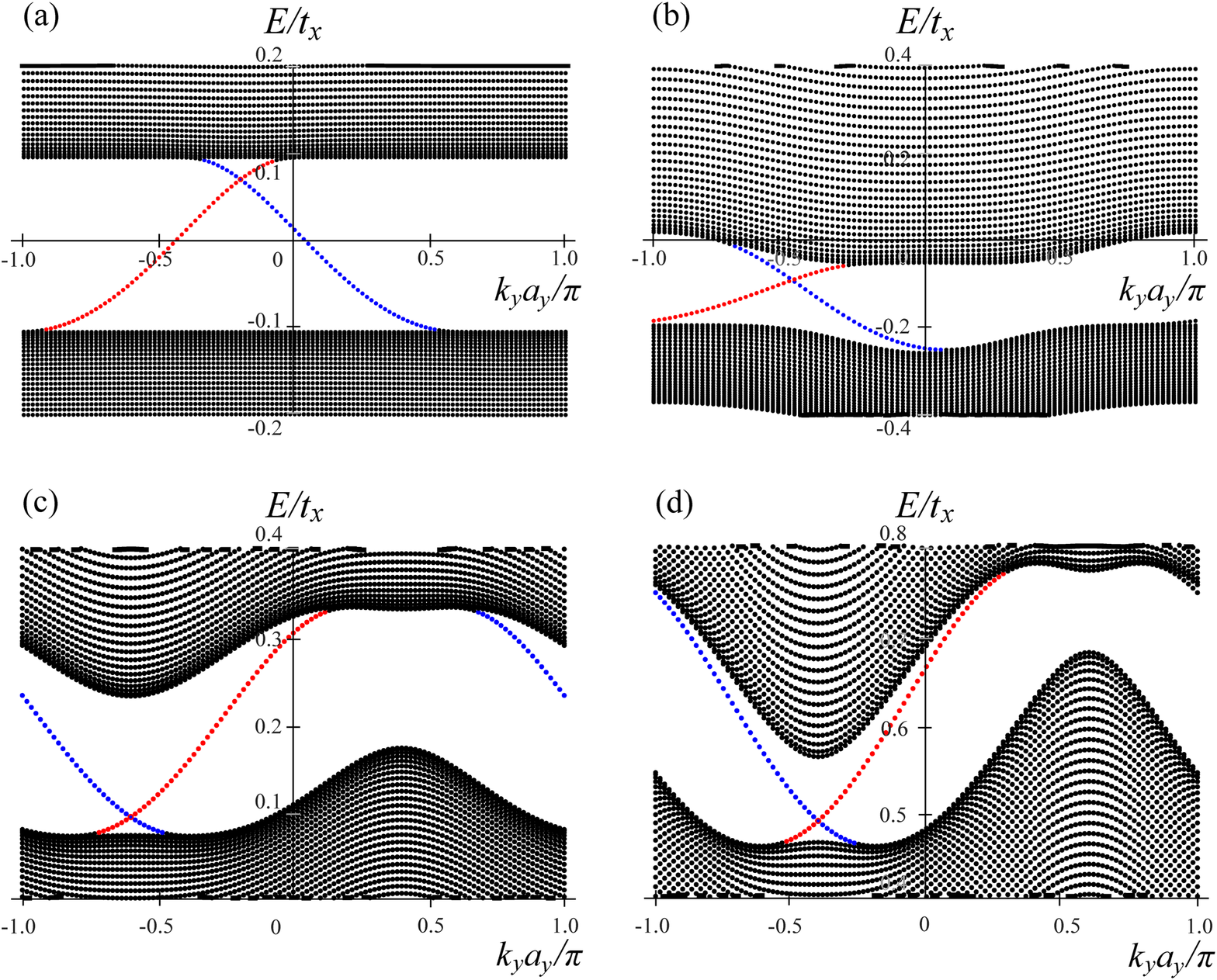}
\caption{(Color online)
Plots of the magnonic band structure, rescaled energy $E/t_x$, for an anisotropic case $J_x \not=J_y$ as function of the rescaled wavevector $k_y a_y/\pi$ 
obtained by numerically solving the tight-binding model [Eq. (\ref{TightBinding2})] for the values
$\theta _0=2\pi/5$, $t_y=0.1 t_x$, showing the first and second Landau levels. 
The periodicity of the vector potential ${\mathbf{A}}_{\rm{m}q}^{\prime} = ({\mathcal{E}} R_q/c) (0, \{x/R_q \}, 0) $ is 
(a) $ q \gg  1$, (b) $q = 6$, (c) $q = 4$, and (d) $q = 3$.
(a) and (b): There is a well-developed gap with a corresponding edge state.
(c) and (d):  There are well-defined edge modes for fixed values of $k_y$, but they coexist with extended bulk modes at different momenta as in the isotropic case $t_x =t_y$.
\label{fig:Edge_numcalSum} }
\end{center}
\end{figure}

%%%%%%%%%%%%%%%%%%%%%%%%%%% 
\section{Magnon spectrum in anisotropic case}
\label{sec:aniso}
%%%%%%%%%%%%%%%%%%%%%%%%%%%
In this Appendix, focusing on the anisotropic case $J_x \not=J_y$, we consider the spectrum around the lowest Landau level, see Fig. \ref{fig:Edge_numcalSum}.
For large values of $q$ such that $ l_{\rm{{\mathcal{E}}}} \ll qa_x$ (see the main text), there is essentially no deviation from the standard case of a uniform gradient, see Fig. \ref{fig:Edge_numcalSum} (a). 
The bulk gap (between the first and second Landau level) stays flat and there is one chiral edge state propagating along $y$ direction at each edge of the strip. 
As  $q$ gets smaller, the bulk spectrum is modified, see Fig. \ref{fig:Edge_numcalSum} (b). 
In particular, the bulk gap is not flat anymore. For even smaller values of $q$, the modification is  very drastic in the sense that the bulk gap is closed \cite{WeylMagnon} and the system becomes gapless; see Figs. \ref{fig:Edge_numcalSum} (c) and (d).
However, like in the isotropic case discussed in the main text, for each given value of $k_y$, there is still a gap in the spectrum and the chiral edge states still exist. 
In comparison to the isotropic case, the anisotropy keeps the Landau bands more flat. 
Thus, a periodically extended flux is best implemented for strongly anisotropic systems.
Also, the sawtooth form of the periodic extension is important; for instance, for a triangular shape the QHE disappears completely.

%%%%%%%%%%%%%%%%%%%%%%%%%%%%%%%%%%%%%%
\section{Magnon quantum Hall conductance}
\label{sec:BulkMagnonConductivity} 
%%%%%%%%%%%%%%%%%%%%%%%%%%%%%%%%%%%%%%
In this Appendix, we provide details on the derivation of the magnon quantum Hall conductance.
Our derivation follows that of Ref. [\onlinecite{Kohmoto}], with the difference that our low-energy excitations are bosonic and given by magnons which we treat in Landau quantization (see also main text). 
The Kubo formula for the Hall conductance \cite{NiuBerry,Kohmoto,TKNN} reads 
\begin{eqnarray}
G ^{y x} &= &  -  i\hbar    \frac{(g\mu _{\rm{B}})^2}{L^2} \sum_{{\mathbf{k}} } \sum_{n } n_{\rm{B}} (E_{n {\mathbf{k}}}) \nonumber  \\
& \times& \sum_{m (\not= n) }    
    \Big[ \frac{\langle  n  {\mathbf{k}}   \mid  v_y  \mid  m  {\mathbf{k}}   \rangle  
                 \langle  m {\mathbf{k}}  \mid  v_x   \mid  n   {\mathbf{k}}  \rangle}{(E_{n {\mathbf{k}}}-E_{m {\mathbf{k}}})^2} - {\rm{H. c.}} \Big],   \nonumber  \\
\label{QHEcal7}  
\end{eqnarray}
where $L^2$ is the area of the system and $ \mid u_{{n}{\mathbf{k}}} \rangle    \equiv   \mid  n  {\mathbf{k}}   \rangle    $ for simplicity.
Since we consider a clean (i.e., disorder-free) bulk, the system is characterized by $\{ \mid  n  {\mathbf{k}}   \rangle  \} $. 
One can easily see the relation $G ^{y x}   =  - G ^{x y} $ from Eq. (\ref{QHEcal7}).
Using the Berry curvature defined by
\begin{eqnarray}
{\Omega} _{n, {\cal{\chi }} }(\mathbf{k})   &\equiv &  i \epsilon _{\chi  \varrho  \tau   }   \sum_{m(\not= n)}
\frac{ \langle  n  {\mathbf{k}}  \mid   \frac{\partial  {\cal{H}}_{{\mathbf{k}} }}{\partial  k_{\varrho}}    \mid     m   {\mathbf{k}}  \rangle   
  \langle  m  {\mathbf{k}}  \mid     \frac{\partial  {\cal{H}}_{{\mathbf{k}} }}{\partial  k_{\tau   }}    \mid      n  {\mathbf{k}}  \rangle  }{(E_{n{\mathbf{k}}} - E_{m{\mathbf{k}}})^2},
  \nonumber     \\
\label{curvature}   
\end{eqnarray}
with  Levi-Civita symbol $\epsilon _{\chi  \varrho  \tau }$ $(\chi, \varrho,  \tau  \in \{x, y, z\})$,
Eq. (\ref{QHEcal7}) can be rewritten as
\begin{eqnarray}
  G ^{y x} =  \frac{(g \mu _{\rm{B}})^2}{{h}} \sum_n  \int_{{\rm{BZ}}}  
({d^2 k}/{2\pi}) n_{\rm{B}} (E_{n {\mathbf{k}}})  {\Omega} _{n, z}(\mathbf{k})   .
\end{eqnarray}

At low temperature $k_{{\rm{B}}} T  \ll  \hbar \omega _c$, only the lowest Landau level $ n=0$ becomes relevant and the almost flat band gives the quantized Hall conductance (see  main text),
$G ^{y x}   \approx    [{(g \mu _{\rm{B}})^2}/{h}] n_{\rm{B}} (E_0^{\ast })  \cdot   \nu _0$.
We thus reach the conclusion that in the clean bulk of two-dimensional insulating magnetic lattices, the magnon Hall conductance in Landau quantization becomes discrete in units of 
\begin{eqnarray}
\frac{{(g \mu _{\rm{B}})^2}}{{h}} n_{\rm{B}} (E_0^{\ast }).
\label{eqn:Our Unit}
\end{eqnarray}
This distinguishes our result from a disordered chiral spin liquid in Ref. [\onlinecite{Haldane2}] where the existence of a gap is assumed; 
in such a disordered system, the Bloch wavevector is no longer a good quantum number since the translation symmetry is broken due to random impurities. However, the role can be instead played by the phase parameters for the boundary condition (phase twist), and  Haldane and Arovas \cite{Haldane2} indeed showed that even in that case, the magnon Hall conductance is still characterized by the Chern number; the bulk-edge correspondence ensures that as long as chiral edge magnon states exist, the quantization of the magnon Hall conductance is robust against disorder effects, which is consistent with Ref. [\onlinecite{RSdisorder}] numerically demonstrating a disordered quantum Hall regime in systems with dipolar interactions.
%{\color{red} noninteracting magnon ??} 
%the Bloch wavevector is no longer a good quantum number due to disordered effects ({\it{e.g.}}, impurities) and the role is instead played by the phase parameters for the boundary condition (phase twist), as in Ref. [\onlinecite{NiuPhaseTwist}].

%
Our result Eq. (\ref{eqn:Our Unit}) shows  that the prefactor of the Chern number depends on temperature and the typical energy value for the almost flat band characterized by the Landau gap (see  main text).
This arises from the intrinsic bosonic properties of magnons that the distribution function cannot be replaced by the Heaviside step function even at zero temperature.
Therefore, the Bose-distribution function $n_{\rm{B}}(E_0^{\ast })$ plays a crucial role in the magnonic QHE.

We note that in sharp contrast to the electric QHE, the magnetic Hall conductance of bulk magnons $G ^{y x} $ 
does not reduce to the form analogous to the St$\check{\rm{r}}$eda formula \cite{Streda} (see Ref. [\onlinecite{TITSC}] for details), 
since the driving force is not the magnetic field $B$ but its gradient $\partial _x B$;
the Hall current density of bulk magnons can be written as 
 $j_i  \equiv  G _{\rm{H}} \epsilon _{ij} \partial _j B  $  with $G _{\rm{H}} =  G ^{y x}$, 
and it becomes $ \partial _i  j_i =  G _{\rm{H}} \epsilon _{ij} \partial _i   \partial _j B =0 $.

Lastly, we mention that the mathematical structure of the magnetic system characterized by the electric vector potential  ${\mathbf{A}}_{\rm{m}} $ in the AC effect  is identical to that of the electronic system \cite{Kohmoto,NiuBerry} by the magnetic vector potential  in the AB effect.
Therefore even without the periodic electric vector potential, the Bloch wavevector still remains a good quantum number to describe the system due to the periodic lattice potential (see Refs. [\onlinecite{Kohmoto,NiuBerry}] for details);
in analogy with the magnetic translation operators \cite{NiuBand} as in Ref. [\onlinecite{NiuBerry}], 
as long as the components of ${\mathbf{A}}_{\rm{m}} $ are linear in $x$ and $y$,
the translation operators \cite{Kohmoto} for enlarged Bravais lattice vector, which commute with the Hamiltonian as well as with each other, can be defined in the electric field gradient.
Then the resulting Bloch wavevector \cite{KevinHallEffect} remains a good quantum number to describe the state in clean systems.
Thus the simultaneous eigenfunction is well characterized by the Landau level index and the Bloch wavevector in the corresponding BZ analogous to the magnetic \cite{NiuBerry,Kohmoto,Zak} BZ for electronic systems.

%%%%%%%%%%%%%%%%%%%%%%%%%%%%%%%%%%%%%%%%%%%%%%%
%%%%%%%%%%%%%%%%%%%%%%%%%%%%%%%%%%%%%%%%%%%%%%%
\begin{table*}
\caption{
\label{tab:table1}
Landau levels of various magnons.
The energy level of nonrelativistic-like magnon is quantized in units of the Landau level index $n$ and the energy level spacing becomes uniform,
while that of relativistic-like magnons is quantized in units of $\sqrt{n}$ and the energy level spacing depends on  $n$.
}
%%%%%%%%%%%%
\begin{ruledtabular}
\begin{tabular}{ccccccc}
%%%%%%%%%%%%%%%%%%%%%%%%%%%%%%%%%%%%%%%%%%%%
   & Nonrelativistic-like magnon & Ferromagnetic Dirac magnon  & Antiferromagnetic Dirac magnon
\\ \hline
  Dispersion: & Quadratic  & Linear &  Linear  \\
  %%%%%%%%%%%%%%%%%%%%%%%%%%%%%%%%%%%%%%%%%%%%
 Hamiltonian: & ${\cal{H}}_{\rm{m}} =    \frac{1}{2m} \big({\mathbf{p}}  +  \frac{g \mu _{\rm{B}}}{c} {\mathbf{A}}_{\rm{m}}   \big)^2$.  
& $ {\cal{H}}_{\rm{{\cal{D}}(F)}} =   v_J {\mathbf{\sigma}} \cdot  \big({\mathbf{p}}  +  \frac{g \mu _{\rm{B}}}{c} {\mathbf{A}}_{\rm{m}} \big) 
+ \hat\varepsilon $.
& $  {\cal{H}}_{\rm{{\cal{D}}(AF)}} =   \sqrt{2} v_J {\mathbf{\sigma}} \cdot  \big({\mathbf{p}}  +  \frac{g \mu _{\rm{B}}}{c} {\mathbf{A}}_{\rm{m}} \big) + g \mu _{\rm{B}} B_0$. \\
%%%%%%%%%%%%%%%%%%%%%%%%%%%%%%%%%%%%%%%%%%%%% 
Landau level: & $E_n   = \hbar \omega _c (n +  1/2) $. 
& $  E_{n \pm }^{\rm{{\cal{D}}(F)}}   = \pm  \hbar \omega _c^{\rm{{\cal{D}}(F)}} \sqrt{n}  + \varepsilon $.  
& $E_{n \pm }^{\rm{{\cal{D}}(AF)}}   = \pm  \hbar \omega _c^{\rm{{\cal{D}}(AF)}} \sqrt{n}  +  g \mu _{\rm{B}} B_0 $.    \\
%%%%%%%%%%%%%%%%%%%%%%%%%%%%%%%%%%%%%%%%%%%%%
 Frequency: & $\omega _c = \frac{g \mu _{\rm{B}}{\mathcal{E}}}{m c^2} \propto {1}/{{ l_{\rm{{\mathcal{E}}}}}^2} $.
& $\omega _c^{\rm{{\cal{D}}(F)}}  = \sqrt{2} v_J/ l_{\rm{{\mathcal{E}}}}\propto {1}/{{ l_{\rm{{\mathcal{E}}}}}}$.  
& $ \omega _c^{\rm{{\cal{D}}(AF)}}  = \sqrt{2} \omega _c^{\rm{{\cal{D}}(F)}} =  2 v_J/ l_{\rm{{\mathcal{E}}}} \propto {1}/{{ l_{\rm{{\mathcal{E}}}}}}$.  \\
%%%%%%%%%%%%%%%%%%%%%%%%%%%%%%%%%%%%%%%%%%%%%
$ E_n - E_{n-1} $: & $E_n - E_{n-1}  = \hbar \omega _c  $. 
& $ E_{n +}^{\rm{{\cal{D}}(F)}} -  E_{n-1 +}^{\rm{{\cal{D}}(F)}}=  \hbar \omega _c^{\rm{{\cal{D}}(F)}} (\sqrt{n} - \sqrt{n-1})$.  
& $E_{n +}^{\rm{{\cal{D}}(AF)}} -  E_{n-1 +}^{\rm{{\cal{D}}(AF)}}=  \hbar \omega _c^{\rm{{\cal{D}}(AF)}} (\sqrt{n} - \sqrt{n-1}) $.    \\
\end{tabular}
\end{ruledtabular}
\end{table*}
%%%%%%%%%%%%%%%%%%%%%%%%%%%%%%%%%%
%%%%%%%%%%%%%%%%%%%%%%%%%%%%%%%%%%

%%%%%%%%%%%%%%%%%%%%%%%%%%%
\section{Landau levels of Dirac magnons}
\label{sec:DM}
%%%%%%%%%%%%%%%%%%%%%%%%%%%
In this Appendix, we consider a relativistic-like magnon with a linear dispersion, namely, {\it{Dirac magnon}} \cite{DiracMagnon},
and clarify the difference from the nonrelativistic-like magnon with the quadratic dispersion (see the main text) in terms of Landau quantization. 

Recently, Fransson {\it{et al.}} \cite{DiracMagnon} have pointed out the possibility that a Dirac-like \cite{Graphene} magnon spectrum is generated on two-dimensional honeycomb lattices (see Ref. [\onlinecite{DiracMagnon}] for details).
Such magnetic excitations with a linear dispersion are called Dirac magnons, and they are robust \cite{DiracMagnon} against magnon-magnon interactions.
Dirac magnons can emerge naturally from the bipartite lattice structure inherent to the honeycomb lattices.

%%%%%%%%%%%%%%%%%%%%%%%%%%%
\subsection{Ferromagnetic Dirac magnon}
\label{subsec:DMferro}
%%%%%%%%%%%%%%%%%%%%%%%%%%%
Around the (so-called) $K$- and $K^{\prime}$-points \cite{DiracMagnon,Graphene} on a ferromagnetic honeycomb spin lattice,
ferromagnetic Dirac magnons in the  presence of an AC phase are described \cite{DiracMagnon,Graphene,magnon2,Mignani} by the Hamiltonian ${\cal{H}}_{\rm{{\cal{D}}(F)}}$,
\begin{eqnarray}
 {\cal{H}}_{\rm{{\cal{D}}(F)}}  =   v_J {\mathbf{\sigma}} \cdot  \Big({\mathbf{p}}  
+  \frac{g \mu _{\rm{B}}}{c} {\mathbf{A}}_{\rm{m}} \Big) 
+ \hat\varepsilon, 
\label{HamiltonianDMLL} 
\end{eqnarray}
where  $\hat\varepsilon = {\rm{diag}}\{ \varepsilon,  \varepsilon  \}$ with $\varepsilon = 3JS + g \mu _{\rm{B}} B_0$, Pauli matrices ${\mathbf{\sigma}}$, 
and $v_J = 3aJS/2\hbar $ is the velocity of Dirac magnons.
The spin length on each sublattice is identical, and it is denoted by $S$. 
%%%%%%%%%%%%%%%%%%%%%%%%%%%%%%%%%%%%%%%%%%%%%%%%%
Using the same procedure as in Appendix \ref{sec:MagnonicLLcal}, the Hamiltonian can be rewritten as
\begin{subequations}
\begin{eqnarray}
%%%%%%%%%%%%%%%%%%%%%%%%%%%%%
{\cal{H}}_{\rm{{\cal{D}}(F)}}
%%%%%%%%%%%%%%%%%%%%%%%%%%%%%
&=&
%%%%%%%%%%%%%%%%%%%%%%%%%%%%%
\begin{pmatrix}
 \varepsilon  & \hbar  \omega _c^{\rm{{\cal{D}}(F)}} a^{\dagger }    \\    \hbar  \omega _c^{\rm{{\cal{D}}(F)}} a   &  \varepsilon
\end{pmatrix}, \\
\label{eqn:HamiltonianDMmatrix}
%%%%%%%%%%%%%%%%%%%%%%%%%%
\omega _c^{\rm{{\cal{D}}(F)}}  & \equiv & \sqrt{2} v_J/ l_{\rm{{\mathcal{E}}}}.
\end{eqnarray}
\end{subequations}
The eigenstate is then given by
\begin{subequations}
\begin{eqnarray}
 \mid   n    \   \rangle   \rangle   &\equiv &
\begin{pmatrix}
 \kappa _n  \mid   n   \rangle      \\   \lambda _n  \mid   n-1   \rangle      
\end{pmatrix}
\    \     {\text{for}}    \       \      n \geq 1,
%%%%%%%%%%%%%%%%%%%%%%%%%%%%%%%%%%
\label{eqn:HamiltonianDMmatrix2}
\\
 \mid   0    \   \rangle   \rangle   &\equiv &
\begin{pmatrix}
       \mid   0   \rangle      \\    0
\end{pmatrix}
\    \     {\text{for}}    \       \      n =0,
\label{eqn:HamiltonianDMmatrix3}
\end{eqnarray}
\end{subequations}
where $ a    \mid   n   \rangle  = \sqrt{n}  \mid   n-1   \rangle$, 
$ a^{\dagger }    \mid   n   \rangle  = \sqrt{n+1}  \mid   n+1   \rangle$,
 $ a   \mid   0   \rangle  = 0$,
and coefficients $ \kappa _n$ and $ \lambda _n$.
%%%%%%%%%%%%%%%%%%%%%%%%%%%%%%
Finally, the eigenvalue equation,
${\cal{H}}_{\rm{{\cal{D}}(F)}} \mid   n    \   \rangle   \rangle = E_n^{\rm{{\cal{D}}(F)}} \mid   n    \   \rangle   \rangle  $,
provides the Landau level for ferromagnetic Dirac magnons
\begin{eqnarray}
E_{n \pm }^{\rm{{\cal{D}}(F)}}   = \pm  \hbar \omega _c^{\rm{{\cal{D}}(F)}} \sqrt{n}  + \varepsilon   \    \    \       {\rm{for}}     \     \  n \in  {\mathbb{N}}_0.
\label{DM_LL} 
\end{eqnarray}
%%%%%%%%%%%%%%%
This can be easily seen as follows \cite{Graphene}; the eigenvalue equation for  $n=0$, 
${\cal{H}}_{\rm{{\cal{D}}(F)}} \mid   0    \   \rangle   \rangle = E_0^{\rm{{\cal{D}}(F)}} \mid   0    \   \rangle   \rangle  $,
gives 
\begin{eqnarray}
E_0^{\rm{{\cal{D}}(F)}} =  \varepsilon.
\label{eqn:FDMcal0}
\end{eqnarray}
The equation for $n \geq 1 $,
 ${\cal{H}}_{\rm{{\cal{D}}(F)}} \mid   n   \   \rangle   \rangle = E_n^{\rm{{\cal{D}}(F)}} \mid   n    \   \rangle   \rangle  $,
provides
\begin{eqnarray}
\begin{pmatrix}
 (\varepsilon \kappa _n +  \hbar  \omega _c^{\rm{{\cal{D}}(F)}} \sqrt{n} \lambda _n ) \mid   n   \rangle      \\    
(\hbar  \omega _c^{\rm{{\cal{D}}(F)}} \sqrt{n}  \kappa _n   + \varepsilon  \lambda _n)  \mid   n-1   \rangle      
\end{pmatrix}
=
\begin{pmatrix}
 E_n^{\rm{{\cal{D}}(F)}} \kappa _n    \mid   n   \rangle      \\    
 E_n^{\rm{{\cal{D}}(F)}} \lambda _n    \mid   n-1   \rangle      
\end{pmatrix}.
 \nonumber  \\
\label{eqn:FDMcal}
\end{eqnarray}
This gives
\begin{subequations}
\begin{eqnarray}
\varepsilon \kappa _n +  \hbar  \omega _c^{\rm{{\cal{D}}(F)}} \sqrt{n} \lambda _n  &=& E_n^{\rm{{\cal{D}}(F)}} \kappa _n,   \\
\hbar  \omega _c^{\rm{{\cal{D}}(F)}} \sqrt{n} \kappa _n   + \varepsilon  \lambda _n &=&  E_n^{\rm{{\cal{D}}(F)}} \lambda _n, 
\label{eqn:FDMcal2}
\end{eqnarray}
\end{subequations}
which can be rewritten as
\begin{eqnarray}
\begin{pmatrix}
 \varepsilon  -  E_n^{\rm{{\cal{D}}(F)}}  & 
\hbar  \omega _c^{\rm{{\cal{D}}(F)}} \sqrt{n}   \\   
 \hbar  \omega _c^{\rm{{\cal{D}}(F)}}  \sqrt{n}   & 
\varepsilon  -  E_n^{\rm{{\cal{D}}(F)}}  
\end{pmatrix}
%%%%%%%%%%%%%%%%%%%%%%%%%%%%%%%
\begin{pmatrix}
 \kappa _n       \\    
 \lambda _n       
\end{pmatrix}
=0.
\label{eqn:FDM7}
\end{eqnarray}
The non-trivial solution reads,
$(\varepsilon  -  E_n^{\rm{{\cal{D}}(F)}})^2- (\hbar  \omega _c^{\rm{{\cal{D}}(F)}}  \sqrt{n})^2=0$,
which provides
\begin{eqnarray}
E_{n \pm }^{\rm{{\cal{D}}(F)}}   = \pm  \hbar \omega _c^{\rm{{\cal{D}}(F)}} \sqrt{n}  + \varepsilon   \    \    \       {\rm{for}}     \     \   n \in  {\mathbb{N}}_+.
\label{DM_LLcal} 
\end{eqnarray}
Finally, Eqs. (\ref{eqn:FDMcal0}) and (\ref{DM_LLcal}) result in the Landau levels for ferromagnetic Dirac magnons [Eq. (\ref{DM_LL})]. 

Like for nonrelativistic-like magnon case, the energy eigenvalue becomes discrete (i.e., Landau level quantization).
However, the energy level spacing of ferromagnetic Dirac magnons, relativistic-like magnons, is not uniform and does depend on the Landau level index $n$,
\begin{eqnarray}
E_{n +}^{\rm{{\cal{D}}(F)}} -  E_{n-1 +}^{\rm{{\cal{D}}(F)}}=  \hbar \omega _c^{\rm{{\cal{D}}(F)}} (\sqrt{n} - \sqrt{n-1})
\       {\rm{for}}   \    n \in  {\mathbb{N}}_+.    \nonumber   \\
\label{DM_LL2} 
\end{eqnarray}
This results from the properties that the energy level for Dirac magnons is quantized in units of $\sqrt{n}$ [see Eq. (\ref{DM_LL})].
These stands in sharp contrast to the nonrelativistic magnons.

%%%%%%%%%%%%%%%%%%%%%%%%%%%
\subsection{Antiferromagnetic Dirac magnon}
\label{subsec:DMAF}
%%%%%%%%%%%%%%%%%%%%%%%%%%%
The Landau level for Dirac magnons on antiferromagnetic honeycomb spin lattices can be derived in the same way.
Around the (so-called) $\Gamma $-point \cite{DiracMagnon,Graphene} on an antiferromagnetic honeycomb spin lattice, 
antiferromagnetic Dirac magnons in the presence  of an AC phase are described \cite{DiracMagnon,Graphene,magnon2,Mignani} by the Hamiltonian ${\cal{H}}_{\rm{{\cal{D}}(AF)}}$,
\begin{eqnarray}
 {\cal{H}}_{\rm{{\cal{D}}(AF)}} =   \sqrt{2} v_J {\mathbf{\sigma}} \cdot  \Big({\mathbf{p}}  
+  \frac{g \mu _{\rm{B}}}{c} {\mathbf{A}}_{\rm{m}}  \Big) + g \mu _{\rm{B}} B_0.  
\label{HamiltonianDMLLAF} 
\end{eqnarray}
The correspondence with the ferromagnetic Dirac magnon [Eq. (\ref{HamiltonianDMLL})] is straightforward.
Replacing $v_J$ by $ \sqrt{2} v_J $ and  $\varepsilon $ by $ g \mu _{\rm{B}} B_0$,
the Landau level for antiferromagnetic Dirac magnons is given by ($ n \in  {\mathbb{N}}_0 $)
\begin{subequations}
\begin{eqnarray}
E_{n \pm }^{\rm{{\cal{D}}(AF)}}   &=& \pm  \hbar \omega _c^{\rm{{\cal{D}}(AF)}} \sqrt{n}  +  g \mu _{\rm{B}} B_0,   \\
\label{AFDM_LL} 
\omega _c^{\rm{{\cal{D}}(AF)}}   &=& 2 v_J/ l_{\rm{{\mathcal{E}}}}.
\end{eqnarray}
\end{subequations}
The frequency of antiferromagnetic Dirac magnons becomes larger by $\sqrt{2}$ times than that of ferromagnetic ones
$\omega _c^{\rm{{\cal{D}}(AF)}}/\omega _c^{\rm{{\cal{D}}(F)}}  = \sqrt{2}$.
Like for ferromagnetic Dirac magnons, 
the energy level for antiferromagnetic Dirac magnons is quantized also in units of $\sqrt{n}$,
and the energy level spacing again depend on the Landau level index $n$,
\begin{eqnarray}
E_{n +}^{\rm{{\cal{D}}(AF)}} -  E_{n-1 +}^{\rm{{\cal{D}}(AF)}}=  \hbar \omega _c^{\rm{{\cal{D}}(AF)}} (\sqrt{n} - \sqrt{n-1})
\       {\rm{for}}   \        n \in  {\mathbb{N}}_+.   \nonumber  \\
\label{DM_LL2AF} 
\end{eqnarray}
The results of Landau levels for nonrelativistic-like and relativistic-like magnons are summarized in Table \ref{tab:table1}.

It is well-known that Dirac fermions in graphene generate an unconventional \cite{QHErelEl,GrapheneQHE} integer QHE due to a quantum anomaly of the lowest Landau level. Therefore, as an outlook, we mention that seeking a possibility for a fractional \cite{KaneFisher} magnonic QHE, it would be interesting to explore the QHE and the WF law of Dirac magnons in Landau quantization and investigate the difference from the ones of nonrelativistic-like magnons.

\bibliography{PumpingRef}

\begin{thebibliography}{90}
\bibitem[{NOT({\natexlab{b}})}]{NOTEmmint}
\bibinfo{note}{Microscopically, magnon-magnon interactions arise from
  Heisenberg models with anisotropic exchange in spin space, see Refs.
  [\onlinecite{KKPD,KPD}] for details where it has been shown even with taking
  into account the next leading order.}
\end{thebibliography}

\end{document}